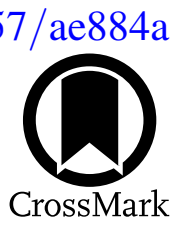

# Black Hole–Galaxy Correlations in Cluster Zoomed-in Simulations: GIZMO-SIMBA and TNG-Cluster

Hangxin Pu[1], Antonios Katsianis[1], Weiguang Cui[2], Romeel Davé[3], Massimo Gaspari[4], Weishan Zhu[1], Xiaohu Yang[5], Jinyi Shangguan[6], Qingshan Wang[1], Ying Tang[1], and Yuchang Li[1]
[1] Sun Yat-Sen University, School of Physics and Astronomy, People's Republic of China; katsianis@mail.sysu.edu.cn
[2] Universidad Autónoma de Madrid, Departamento de Física Teórica, Spain; weiguang.cui@uam.es
[3] University of Edinburgh, UK
[4] University of Modena and Reggio Emilia, Department of Physics, Italy
[5] Shanghai Jiao Tong University, Tsung-Dao Lee Institute & School of Physics and Astronomy, People's Republic of China
[6] Peking University, The Kavli Institute for Astronomy and Astrophysics, People's Republic of China

## Abstract

We investigate the coevolution of supermassive black holes (SMBHs) and central galaxies in massive clusters using the GIZMO-SIMBA and TNG-Cluster zoomed-in simulations at $z = 0$–5. We find that the distinct subgrid physics of these two models suggest fundamentally different evolutionary pathways. On the one hand, GIZMO-SIMBA employs torque-limited accretion and predicts a supply-driven scenario where the SMBHs rapidly assemble synchronized with dark matter halo ($M_{200c}$) growth (i.e., the halo mass–BH mass relation is set by $z = 3.0$ and similar to the present-day relationship). On the other hand, TNG-Cluster, exhibits a feedback-regulated growth phase delayed by an early thermal suppression. While both models successfully reproduce some local black hole–galaxy scaling relations, they imply significantly different evolution for these relations. Analysis of the BH mass–gas mass ratio relations suggests that TNG-Cluster's isotropic kinetic winds efficiently deplete cold gas, resulting in a "hard quench" of star formation. In the black hole accretion rate (BHAR)–star formation rate relation, we find that both simulations successfully reproduce the decoupling of BHAR and star formation observed in recent massive cluster ellipticals. The divergent evolutionary trends emphasize the importance of the multiphase intracluster medium; while these subgrid models do not have the necessary resolution and employ distinct formalisms, the sustained BHAR in quenched systems resemble outcomes broadly consistent with modern multiphase feeding paradigms, such as chaotic cold accretion in turbulent cluster cores. Furthermore, we demonstrate that for both models, black hole mass is a primary regulator of atomic and molecular gas depletion in galaxy clusters.

## 1. Introduction

Supermassive black holes (SMBHs) are believed to play an important role in galaxy formation and evolution. Observations across the local Universe have revealed tight empirical correlations between the mass of central SMBHs and the global properties of their host galaxies. Some of these correlations are found with stellar mass, stellar velocity dispersion, and bulge mass (J. Magorrian et al. 1998; K. Gebhardt et al. 2000; D. Merritt & L. Ferrarese 2001; L. Ferrarese 2002; S. Tremaine et al. 2002; N. Häring & H.-W. Rix 2004). These scaling relations suggest a deep physical connection between black hole growth and galaxy assembly, despite the enormous differences between the spatial scales of SMBHs ($\sim 10^{-6}$ pc) and their host galaxies (∼several kiloparsecs). We note, however, that these relationships are possibly subject to intrinsic scatter, morphology dependence, and environmental effects; thus, they may not be universal or time-invariant, as previously thought (J. Kormendy & L. C. Ho 2013; N. J. McConnell & C.-P. Ma 2013; D. Baron & B. Ménard 2019; N. Sahu et al. 2019a, 2019b; W. Ma et al. 2025).

Moving to theory, in the prevailing Λ cold dark matter paradigm (S. D. M. White & M. J. Rees 1978), in order to reproduce key observables (A. Katsianis et al. 2021a; A. Traina et al. 2026), it is required to add an efficient mechanism to regulate star formation in massive halos. Without such regulation, radiative cooling leads to excessive star formation, producing galaxies that are far more massive and more blue than the ones observed. Feedback from active galactic nuclei (AGNs) has been suggested as the leading solution to this long-standing "overcooling problem" for massive halos (J. Silk & M. J. Rees 1998; T. Di Matteo et al. 2005; D. J. Croton et al. 2006; A. V. Kravtsov & S. Borgani 2012; R. S. Somerville & R. Davé 2015; A. Katsianis et al. 2017; T. Naab & J. P. Ostriker 2017). Both theoretical arguments and observations support a picture in which SMBHs inject energy and momentum into their surroundings, suppressing gas cooling, depleting cold-gas reservoirs, and ultimately quenching star formation in massive galaxies (P. N. Best et al. 2007; K. Schawinski et al. 2007; T. M. Heckman & P. N. Best 2014; B. A. Terrazas et al. 2017; W. Cui et al. 2021). However, while the necessity of AGN feedback is currently accepted as one of the most important candidates for quenching high-mass galaxies, the physical mechanisms governing black hole accretion and the coupling of feedback to the interstellar and circumgalactic medium (CGM) remain among the largest uncertainties in galaxy formation theory (F. Yuan et al. 2018).

Cosmological hydrodynamical simulations provide a powerful framework for exploring the coevolution of SMBHs and galaxies within a fully self-consistent cosmological context. Over the past decade, several state-of-the-art simulation suites—such as Illustris, IllustrisTNG, EAGLE, SIMBA and COLIBRE—have achieved remarkable success in reproducing key galaxy observables at low redshift, including stellar mass functions, galaxy colors, and black hole–galaxy scaling relations (M. Vogelsberger et al. 2014; J. Schaye et al. 2015, 2026; R. Davé et al. 2019; N. Thomas et al. 2019; Y. Li et al. 2020; D. Nelson et al. 2019). Despite these successes, different simulations often rely on fundamentally different subgrid prescriptions for black hole growth and AGN feedback. As a result, they frequently predict divergent evolutionary histories for SMBHs, particularly at high redshift and in massive halos (M. Habouzit et al. 2021, 2022).

An important distinction among modern simulations lies in the treatment of black hole accretion. For example, large-volume simulations, like IllustrisTNG, adopt Bondi–Hoyle–Lyttleton–like accretion models (H. Bondi 1952), in which the black hole growth rate is determined by the local gas density, sound speed, and relative velocity, often capped at the Eddington limit (V. Springel et al. 2005; C. M. Booth & J. Schaye 2009; R. Weinberger et al. 2017; A. Pillepich et al. 2018). However, in the context of cosmological simulations, these Bondi-based prescriptions serve primarily as a subgrid proxy and may not fully represent the true SMBH fueling mode, particularly within cluster cores where key physical ingredients—such as turbulence, multiphase condensation, angular-momentum cancellation, and unresolved feeding geometry—are frequently missing. Crucially, modern theoretical frameworks like chaotic cold accretion (CCA) suggest that while accretion can be suppressed significantly below the Bondi rate in hot, rotating flows (M. Gaspari et al. 2015), it can also be boosted far above it through top-down condensation (M. Gaspari et al. 2017). This creates a complex macro-meso-micro feeding/feedback loop that couples the cluster environment to the central engine (M. Gaspari et al. 2020). In the TNG framework, black hole growth remains largely self-regulated by feedback, as AGN heating raises gas entropy to suppress further accretion. In contrast, the SIMBA simulation introduced a physically motivated gravitational torque-limited accretion model for cold gas, in which angular-momentum transport within galactic disks regulates the inflow of gas onto the SMBH (P. F. Hopkins & E. Quataert 2011; D. Anglés-Alcázar et al. 2017). This model naturally links black hole growth to galaxy-scale dynamics and does not require strong self-regulation to reproduce observed scaling relations (D. Anglés-Alcázar et al. 2017; R. Davé et al. 2019; N. Thomas et al. 2019). The two accretion paradigms, therefore, represent fundamentally different physical pictures of SMBH growth and their host galaxies.

The impact of these differences between simulations is expected to be evident in dense environments, particularly in galaxy clusters. In massive halos, the relevant question is not only whether feedback quenches, but whether cooling, turbulence, and uplift seed top-down multiphase condensation that can decouple instantaneous black hole accretion rate (BHAR) from star formation rate (SFR; M. Gaspari et al. 2018; H. Olivares et al. 2019). The brightest cluster galaxies (BCGs) residing at the centers of the most massive dark matter halos are affected by cooling flows, hot intracluster gas, and repeated mergers that create extreme conditions for black hole growth. Observations of clusters reveal tight connections between AGN activity and the thermodynamic state of the intracluster medium (ICM), including X-ray cavities, multiphase gas precipitation, and suppressed cooling rates (A. C. Fabian 2012; M. Gaspari et al. 2012, 2017; B. R. McNamara & P. E. J. Nulsen 2012; F. J. Jennings et al. 2025). Recent studies have further shown that SMBH mass correlates strongly with halo mass in BCGs (W. Cui et al. 2022), suggesting that black hole growth in clusters may be more directly linked to halo-scale processes, an effect that is much weaker for field galaxies (Á. Bogdán et al. 2018; M. Gaspari et al. 2019). These findings can be used as a test to evaluate if the adopted accretion and feedback prescriptions in simulations can simultaneously reproduce both the galaxy-scale and cluster-scale observables.

To robustly assess how black hole accretion and feedback physics shape SMBH–galaxy coevolution in the most massive environments, high-resolution cluster zoomed-in simulations are essential. The Three Hundred project employs the GIZMO code with the SIMBA galaxy formation model to simulate a large sample of galaxy clusters with sufficient resolution to resolve BCGs and their central black holes (W. Cui et al. 2018, 2022). Recently, the TNG-Cluster simulation suite has been introduced as a complementary effort based on the IllustrisTNG model and the AREPO moving-mesh hydrodynamics code (V. Springel 2010), specifically designed to study AGN feedback and cooling flows in massive halos (D. Nelson et al. 2024). Although both suites aim to model similar physical systems, they differ fundamentally in their hydrodynamical solvers, star formation models, black hole accretion prescriptions, and feedback implementations.

In this work, we present a systematic, side-by-side comparison of SMBH growth and black hole–galaxy scaling relations in the GIZMO-SIMBA and TNG-Cluster simulations. Focusing on the SMBHs and their host central galaxies in galaxy clusters, we investigate how differences in accretion physics and feedback geometry propagate into observable relations between black hole mass and halo mass, stellar mass, velocity dispersion, gas fractions, star formation activity, and BHARs across cosmic time. By directly confronting two leading but physically distinct simulation frameworks in the cluster regime, this study aims to clarify the physical origin of SMBH–galaxy correlations and to assess the robustness of these relations as probes of galaxy formation physics for some of the most extreme environments of the Universe.

## 2. Simulations

### 2.1. GIZMO-SIMBA

The GIZMO-SIMBA run is a pioneering hydrodynamical simulation suite performed within The Three Hundred project (W. Cui et al. 2022). It applies the state-of-the-art SIMBA galaxy formation model (R. Davé et al. 2019) implemented in the Meshless Finite Mass (MFM) GIZMO code (P. F. Hopkins 2015) to 324 large-scale cluster environments originally selected from the MultiDark Planck 2 (MDPL2; A. Klypin et al. 2016) dark-matter-only simulation.

While the GIZMO-SIMBA run retains the identical physical modules and feedback parameters as the original SIMBA box ($100\,h^{-1}$ Mpc), it is tailored for cluster-scale studies with specific setup differences. The parent MDPL2 simulation spans a cubic volume of $1\,h^{-1}$ Gpc with $3840^3$ dark matter particles, each of mass $1.5 \times 10^9\,h^{-1}\,M_\odot$. In the GIZMO-SIMBA re-simulations, the high-resolution Lagrangian regions maintain this mass resolution, with gas particles of $m_{\rm gas} = 2.36 \times 10^8\, h^{-1}\, M_\odot$ based on the Planck cosmic baryon fraction (W. Cui et al. 2022). Outside of the selected zoomed-in region of 15 $h^{-1}$ Mpc in radius, the layers of low-resolution dark matter particles are used to represent the cosmological density field. Compared to $m_{\rm gas} = 1.82 \times 10^7\, M_\odot$ in the original SIMBA, the resolution is lower to accommodate the vast volume. Despite these resolution differences, the subgrid parameters in GIZMO-SIMBA were recalibrated to match some specific $z = 0$ global cluster stellar properties, namely, the stellar mass fraction within $R_{500}$, the BCG stellar mass–halo mass relation, and the satellite stellar mass function (W. Cui et al. 2022). Crucially, black hole masses and black hole–galaxy scaling relations were not part of this calibration process, and it was actually done with only one cluster. Therefore, the broad consistency between the GIZMO-SIMBA black hole scaling relations and observations reported in this work represents an independent prediction of the model rather than a direct consequence of the recalibration. More details on the simulation setup and parameter changes with respect to the original SIMBA can be found in R. Davé et al. (2019) and W. Cui et al. (2022).

#### 2.1.1. Black Hole Seeding and Accretion

GIZMO-SIMBA adopts an accretion model that acknowledges the distinct dynamical behaviors of cold, rotationally supported gas versus hot, pressure-supported gas. Black holes are seeded with a mass of $M_{\rm seed} = 10^5\, h^{-1}\, M_\odot$ into compact stellar distributions ($M_* \gtrsim 3 \times 10^{10}\, h^{-1}\, M_\odot$) that do not already host a BH (W. Cui et al. 2022). This ensures seeds are placed in the centers of resolved galaxies (R. Davé et al. 2019).

The simulation distinguishes between two growth modes based on the temperature of the accreting gas ($T_{\rm gas}$). For cold gas ($T \leqslant 10^5$ K) residing in rotation-supported disks, the inflow is limited by angular-momentum removal rather than gas supply. GIZMO-SIMBA implements the torque-limited accretion (cold mode) model (P. F. Hopkins & E. Quataert 2011; D. Anglés-Alcázar et al. 2013, 2015, 2017) when accreting gas with $T < 10^5$ K, where gravitational instabilities (e.g., bars, spirals) drive nonaxisymmetric torques that effectively transport angular momentum outward. This approach explicitly links BH growth to the dynamical state of the host galaxy rather than just the central gas density. It effectively decouples BH growth from the sole dependency on thermal pressure support found in Bondi models, allowing for efficient accretion even in gas that is not centrally concentrated (M. Habouzit et al. 2022). For $T > 10^5$ K, Bondi-like accretion is adopted here as the hot-mode subgrid prescription. True SMBH fueling in massive halos may differ substantially from the Bondi estimate because of multiphase structure and unresolved angular-momentum transport.

#### 2.1.2. Kinetic and X-Ray Feedback

GIZMO-SIMBA uses kinetic feedback as its main method for AGN feedback. Instead of just adding heat, it injects momentum to mimic the real impact of AGN winds and jets (R. Davé et al. 2019). There are two main modes of feedback based on the black hole's accretion rate. The quasar mode occurs during high accretion and drives fast, multiphase winds that resemble broad absorption line outflows (R. Maiolino et al. 2012; E. Sturm et al. 2011). When accretion drops, the simulation switches to jet mode (P. N. Best & T. M. Heckman 2012; I. Barišić et al. 2017), launching very-high-velocity, collimated jets that are associated with radio-loud galaxies (B. R. McNamara & P. E. J. Nulsen 2007; A. C. Fabian 2012). These jets are able to shock-heat the surrounding gas and effectively suppress cooling in massive clusters (N. Thomas et al. 2019; W. Cui et al. 2022). Finally, an X-ray feedback mode activates under specific conditions to volumetrically heat diffuse gas, acting as a final mechanism to prevent gas cooling down to rejuvenate in quenched galaxies (E. Choi et al. 2012).

### 2.2. TNG-Cluster

TNG-Cluster (D. Nelson et al. 2024) is an extension of the IllustrisTNG suite, explicitly designed to overcome the volume limitations of the original simulations. While the TNG300 box (205 $h^{-1}$ Mpc) provided a statistical sample of groups, it contained only a handful of clusters with $M_{200c} > 10^{14}\, h^{-1}\, M_\odot$ ($M_{200c}$ is defined as the spherical overdensity mass, representing the total mass enclosed within a sphere of radius $R_{200c}$, where the mean internal density is 200 times the critical density of the Universe ($\rho_c$) at the given redshift). TNG-Cluster addresses this by re-simulating 352 massive galaxy clusters selected from a 1 Gpc parent volume, providing survey-scale statistics for the most massive objects in the Universe.

The simulation employs the AREPO code (V. Springel 2010), utilizing a moving unstructured Voronoi mesh. Regarding resolution, TNG-Cluster matches the resolution of TNG300 (the intermediate-resolution level of the suite), with a baryon mass resolution of $m_{\rm bar} \approx 1.2 \times 10^7\, M_\odot$ and dark matter particle mass of $m_{\rm DM} \approx 6.1 \times 10^7\, M_\odot$. A critical design principle of TNG-Cluster is that it keeps the physical model parameters entirely fixed to those used in TNG100 and TNG300 (D. Nelson et al. 2024). This enables seamless comparisons across mass scales. More details on the simulation setup can be found in D. Nelson et al. (2024).

#### 2.2.1. Black Hole Seeding and Accretion

In TNG-Cluster, black hole seeds of mass $M_{\rm seed} = 8 \times 10^5\, h^{-1}\, M_\odot$ are inserted into halos once they reach $M_{\rm halo} \approx 5 \times 10^{10}\, h^{-1}\, M_\odot$ (R. Weinberger et al. 2018). The notation $M_{\rm halo}$ is used to specifically denote the Friends-of-Friends (FOF) halo mass. This mass is defined as the sum of the masses of all particles (dark matter, gas, stars, and black holes) linked together by the FOF algorithm during the simulation run, which does not assume a spherical geometry. Accretion follows the Bondi formalism, but the sound speed includes magnetic pressure contributions, which naturally regulate inflow in hot cluster plasmas without requiring boost factors (R. Pakmor et al. 2011; R. Pakmor & V. Springel 2013; D. Nelson et al. 2024). The radiative efficiency is $\epsilon_r = 0.2$, and accretion is Eddington-limited.

#### 2.2.2. Thermal and Kinetic Feedback

Feedback operates in two modes based on the accretion ratio $\chi = \dot{M}_{\rm Bondi}/\dot{M}_{\rm Edd}$. The threshold is $\chi_{\rm limit} = \min[0.002(M_{\rm BH}/10^8 M_\odot)^2, 0.1]$ (V. Springel et al. 2005; M. Vogelsberger et al. 2013). Above this, thermal quasar mode deposits energy at a rate $\dot{E}_{\rm therm} = 0.02 \dot{M}_{\rm BH} c^2$. Below it, kinetic mode accumulates energy until releasing stochastic, pulsed winds with an efficiency that scales with density as $\epsilon_{f,\rm kin} = \min\left(\frac{\rho}{0.05\rho_{\rm SF}}, 0.2\right)$, ensuring strong feedback in dense cluster cores while preserving the diffuse intergalactic medium (A. Pillepich et al. 2018; R. Weinberger et al. 2018).

**Table 1**
Key Asymmetries between GIZMO-SIMBA and TNG-Cluster

| Feature | GIZMO-SIMBA | TNG-Cluster |
|---|---|---|
| Hydrodynamic Solver | Meshless Finite Mass (MFM; GIZMO) | Moving Voronoi Mesh (AREPO) |
| Baryonic Mass Res. | $m_{\rm gas} = 2.36 \times 10^8\,h^{-1}\,M_\odot$ | $m_{\rm bar} = 1.2 \times 10^7\,M_\odot$ |
| BH Seed Mass | $10^5\,h^{-1}\,M_\odot$ | $8 \times 10^5\,h^{-1}\,M_\odot$ |
| Seeding Criterion | Resolved stellar mass ($M_* \gtrsim 3 \times 10^{10}\,h^{-1}\,M_\odot$) | Halo mass threshold ($M_{\rm halo} \gtrsim 5 \times 10^{10}\,h^{-1}\,M_\odot$) |
| Accretion Model | Torque-limited (cold) + Bondi (hot) | Bondi |
| Feedback Geometry | Radiative mode (high accretion) + jet mode (low accretion) | Thermal quasar mode (high accretion) + kinetic winds (low accretion) |

### 2.3. Fairness of Comparison: Numerical Asymmetries and CAESAR Catalogs

It is important to clarify that in both GIZMO-SIMBA and TNG-Cluster suites, the re-simulated clusters are drawn in a highly consistent manner. Both suites extract their clusters from parent dark-matter-only cosmological boxes of large volumes ($\gtrsim$1 Gpc$^3$). In both suites, the target clusters were selected exclusively based on their halo mass at $z = 0$. A difference that emerges is that the Three Hundred project selected the 324 most massive halos ($M_{\rm vir} \geqslant 8 \times 10^{14}\,h^{-1}\,M_\odot$; W. Cui et al. 2018), while TNG-Cluster selected the 352 halos above $10^{15}\,M_\odot$ and a random selection down to $10^{14.3}\,M_\odot$ to flatten the mass distribution (D. Nelson et al. 2024). Because both samples involve the most massive halos drawn from $\gtrsim$1 Gpc$^3$ volumes purely by $z = 0$ mass thresholds, neither suite imposes secondary selection criteria. They fundamentally probe similar cosmological regimes: meaning the highest-mass overdensities in the cosmic web. In addition, The Three Hundred has a much larger zoomed-in high-resolution region (about 15 h$^{-1}$ Mpc in radius) than TNG-Clusters (3 times larger in volume where these clusters' dark matter particles occupied at the initial condition), we only selected these uncontaminated objects within the high-resolution region for comparison in this study. At lower halo mass, the statistics may be more robust from The Three Hundred simulation. However, the clusters' statistics are comparable between the two.

To ensure a transparent physical interpretation of the results presented in Section 3, it is essential to acknowledge that GIZMO-SIMBA and TNG-Cluster represent two distinct "modeling packages." The differences in their predictions arise from a combination of several numerical and physical factors that are changed simultaneously. These asymmetries are summarized in Table 1. The reader should keep in mind these results as a comparison of two self-consistent but multivariable simulation frameworks rather than a controlled study of a single physical parameter.

To provide a consistent and rigorous comparison of the interstellar medium (ISM) phases across both simulation suites, we applied an identical gas-partitioning model to the outputs of GIZMO-SIMBA and TNG-Cluster. While GIZMO-SIMBA natively tracks the molecular hydrogen ($H_2$) fraction as part of its star formation prescription, TNG-Cluster typically requires post-processing to disentangle the atomic and molecular components from the total gas mass.

To eliminate methodological biases, we processed the TNG-Cluster gas cells using the subgrid prescription described in R. Davé et al. (2019). This model utilizes the local gas pressure and metallicity to calculate the molecular fraction ($f_{\rm H_2}$) based on the equation in M. R. Krumholz & N. Y. Gnedin (2011), with the remaining neutral gas assigned to the atomic (H I) phase. By adopting this uniform analysis pipeline, we ensure that the divergent gas depletion signatures discussed below are driven by the simulations' underlying feedback physics rather than differences in the ISM partitioning methodology.

Moreover, to ensure a consistent and meaningful comparison between the GIZMO-SIMBA and TNG-Cluster simulations, we generate galaxy and halo catalogs using the identical analysis pipeline. Halo properties such as $M_{200c}$ are derived using the AHF (Amiga Halo Finder; S. R. Knollmann & A. Knebe 2009), which employs a spherical overdensity algorithm. Based on the established AHF halo catalog, we utilize the CAESAR package to identify galaxies and compute their detailed physical properties.

CAESAR identifies galaxies using a 6D FoF algorithm within a given halo, which considers both the spatial and velocity distributions of star, dense gas, and black hole particles. This approach ensures that galaxies are identified consistently across different simulation codes, regardless of whether the underlying hydrodynamic solver is smoothed-particle-hydrodynamics-based or MFM-based. For each identified object, CAESAR provides a wide range of pre-computed physical and photometric properties, including stellar masses, SFRs, and black hole properties.

Because this analysis is run on the AHF halo catalog, it is easy to identify the CAESAR galaxies within the corresponding AHF halos. Following (W. Cui et al. 2022), the BCG is defined as the most massive CAESAR galaxy located near the potential minimum of the main cluster halo. In this paper, the terms "BCG" and "central galaxy" are used interchangeably throughout the remainder of this paper with no physical distinction. By applying this identical cataloging and selection procedure to both GIZMO-SIMBA and TNG-Cluster, we ensure that any divergent evolutionary pathways or scaling relations identified in this study are driven by the underlying subgrid physics rather than differences in the analysis methodology. Note that while the volume of each The Three Hundred zoomed-in region is larger than that of TNG-Cluster, we ensure a consistent comparison by extracting all resolved halos within these high-resolution regions and analyzing exclusively their central galaxies. This provides a robust sample spanning a wide dynamic range of halo masses.

When deriving extensive and intensive galaxy properties from the catalogs, we apply specific aperture definitions motivated by physical considerations. For the stellar mass ($M_*$) of the central galaxies, we apply a fixed 3D spherical aperture of 50 physical kpc centered on the potential minimum. This is a standard and robust practice for comparing simulated BCGs, as it effectively captures the stellar main body while minimizing the contribution from the extended intracluster light (ICL). Conversely, the gas masses (total gas, atomic H I, and molecular $H_2$) and the 1D stellar velocity dispersion ($\sigma_*$) are not restricted to this 50 kpc aperture. Instead, they are extracted directly from the CAESAR catalogs, which compute these values globally using a 6D FOF algorithm to assign particles to galaxies by the halo finder. This prevents the artificial spatial truncation of extended cold-gas reservoirs and preserves the dynamical self-consistency of the bounded system. As both simulations are consistently analyzed by CAESAR, from which the quantities are drawn, we think the comparisons are fair between the two simulations.

### 2.4. Data Distribution and Sample Selection Constraints

To provide context for the scaling relations analyzed in this work, Figure 1 presents the overall distributions of fundamental host halo and galaxy properties—namely $M_{200c}$, $M_*$, SFR, and specific star formation rate (sSFR)—across the redshift range $z = 0$–5 for both the GIZMO-SIMBA and TNG-Cluster simulations. Those galaxies with stellar masses near the mass seeding criterion (i.e., $3 \times 10^{10}\, h^{-1}\, M_\odot$ for GIZMO-SIMBA) or BH masses near the same seeding mass are excluded from this distribution and from our study. While the mass distributions ($M_{200c}$ and $M_*$) broadly overlap between the two models, their underlying distributions exhibit distinct morphological differences indicative of their respective subgrid accretion and feedback prescriptions.

A critical consideration arises when analyzing the star formation properties (SFR and sSFR). As illustrated in the third row of Figure 1, to visualize these fully quenched populations alongside the active ones on a logarithmic scale, we have artificially assigned them values of $\log_{10}(\mathrm{SFR}) = -3$ for GIZMO-SIMBA and $\log_{10}(\mathrm{SFR}) = -4$ for TNG-Cluster. Both simulations produce a substantial number of galaxies that are completely quenched, yielding an exactly zero SFR. It is worth highlighting that while these exactly zero-SFR systems form a significant population in both simulations, they are entirely absent from the observational samples, which inherently report measurable nonzero values or observational upper limits. Therefore, it is crucial to note that these exactly zero-SFR data points are strictly excluded from the logarithmic scaling relation plots involving SFR or sSFR in the subsequent analysis. Readers should keep the above in mind when comparing the extreme low-sSFR bounds of the two simulations and observations.

## 3. Results

In this section, we present a systematic comparison of the coevolution of SMBHs and their host galaxies in the GIZMO-SIMBA and TNG-Cluster simulations. Focusing on the extreme environments of galaxy clusters, we aim to disentangle how distinct subgrid prescriptions for accretion and feedback manifest in observable scaling relations.

### 3.1. Black Hole Mass–Halo Mass Relation

Figure 2 presents the evolution of the scaling relation between SMBH mass ($M_{\mathrm{BH}}$) and host halo mass ($M_{200c}$) for central galaxies from $z = 5$ to $z = 0$. To highlight the simulation predictions, we compare them at $z = 0$ against a suite of recent observational data points and fitting relation.

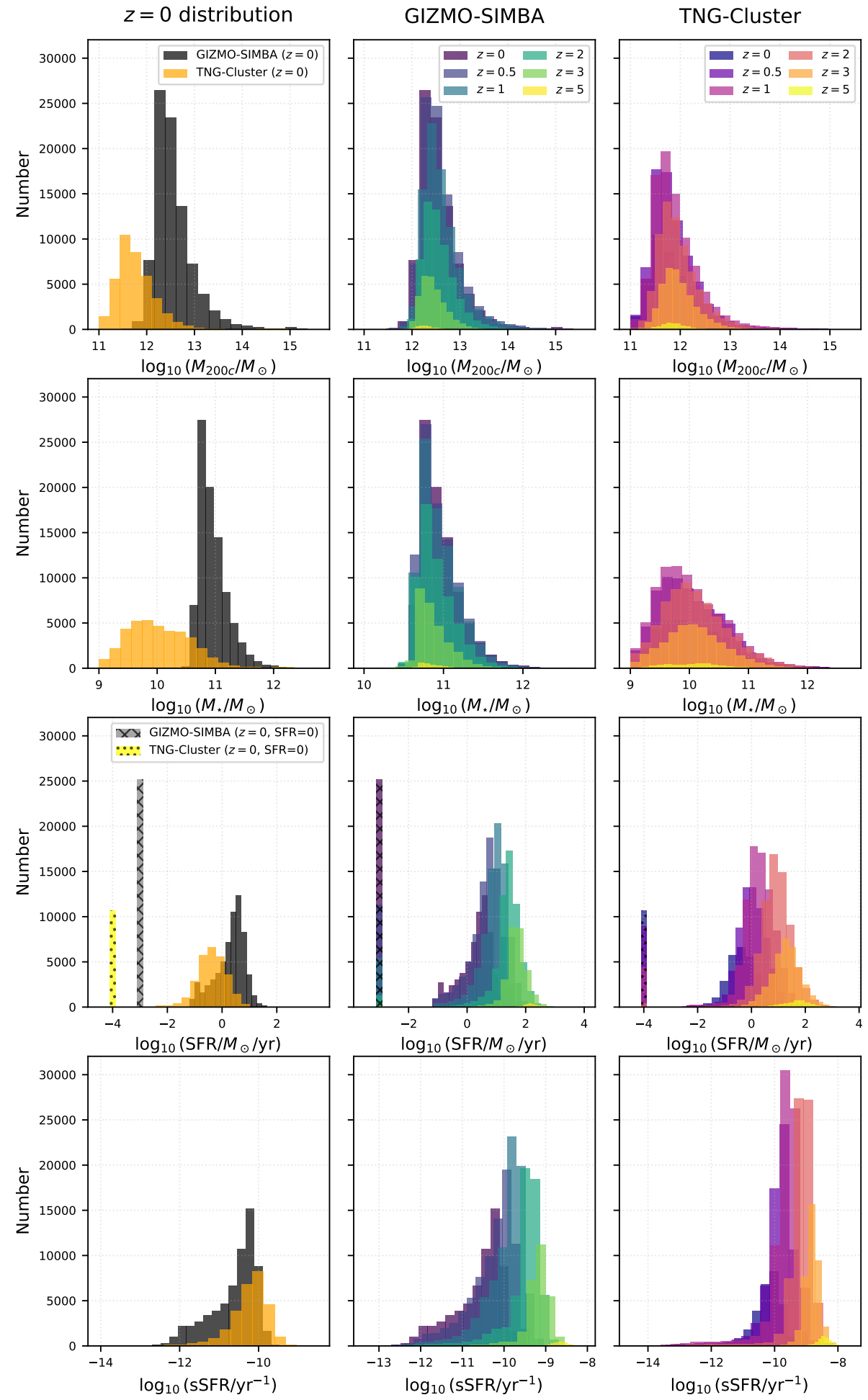

**Figure 1.** The distribution of key halo and galaxy properties for central galaxies in the GIZMO-SIMBA and TNG-Cluster simulations. The left column presents a direct comparison of the distributions at $z = 0$. The middle and right columns show the redshift evolution from $z = 5$ to $z = 0$ for GIZMO-SIMBA and TNG-Cluster, respectively. From top to bottom, the rows display the logarithmic distributions of halo mass ($M_{200c}/M_\odot$), stellar mass ($M_*/M_\odot$), star formation rate ($\mathrm{SFR}/M_\odot\ \mathrm{yr}^{-1}$), and specific star formation rate (sSFR $\mathrm{yr}^{-1}$). For visualization purposes on the logarithmic scale in the SFR panels (third row), galaxies with an exactly zero SFR have been artificially assigned values of $\log_{10}(\mathrm{SFR}) = -3$ in GIZMO-SIMBA (black hatched bar) and $\log_{10}(\mathrm{SFR}) = -4$ in TNG-Cluster (yellow dotted bar).

Both simulations predict a monotonic increase of black hole mass with halo mass, yet they exhibit distinct normalizations and morphological behaviors on the scaling plane. For less-massive halos ($M_{200c} < 10^{13}\, M_\odot$), there are substantive differences in the predictions. GIZMO-SIMBA aligns closely with the M. Gaspari et al. (2019) relation, tracking the lower-mass data points from A. Marasco et al. (2021). TNG-Cluster consistently produces more-massive black holes at fixed $M_{200c}$ but also tracks the higher-mass data points from A. Marasco et al. (2021).

For larger masses ($M_{200c} > 10^{13}\, M_\odot$), the two models are much more similar. Both simulations broadly lie in the range of the data at $M_{200c} = 13$–14, intersecting the Ã. Bogdán et al. (2018) measurements and aligning well with the M. Gaspari et al. (2019) scaling relation and its intrinsic scatter. And GIZMO-SIMBA produces slightly more-massive BH masses compared to TNG-Cluster at $M_{200c} > 10^{14}\,M_{\odot}$, by no more than 2 times. Both simulations agree with the data point from Ã. Bogdán et al. (2018) at this high-mass region.

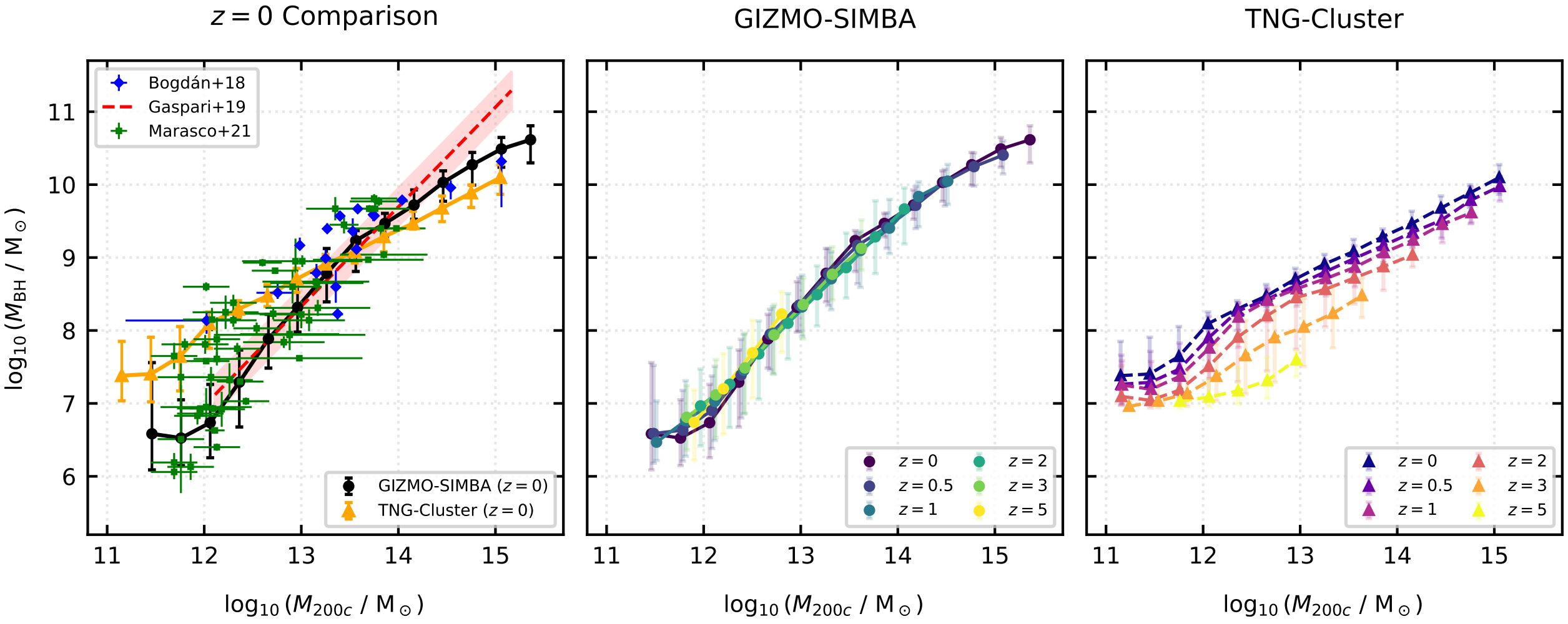

**Figure 2.** The relationship between SMBH mass ($M_{\rm BH}$) and host halo mass ($M_{200c}$) for central galaxies in the GIZMO-SIMBA and TNG-Cluster simulations. The solid and dashed lines connecting the data points represent the median values of the simulated distributions within each mass bin. Error bars denote the 16th–84th percentile range. Left panel: the scaling relation at $z = 0$. GIZMO-SIMBA central galaxies are represented by black circles connected by solid lines, while TNG-Cluster central galaxies are shown as orange triangles connected by solid lines. Observational data points are overlaid for comparison: individual data points from Ã. Bogdán et al. (2018) (blue diamonds) and A. Marasco et al. (2021) (green squares), as well as the fitting scaling relation from M. Gaspari et al. (2019) (red dashed line, with the shaded region representing the intrinsic scatter). Middle and right panels: the redshift evolution of the $M_{\rm BH}$–$M_{200c}$ relation specifically for GIZMO-SIMBA (middle) and TNG-Cluster (right). Data points are color-coded by redshift, tracing the evolutionary track from $z = 5$ to $z = 0$. Error bars represent the standard deviation within each mass bin.

The redshift evolution panels reveal a fundamental difference in the assembly history of the two simulations. GIZMO-SIMBA shows remarkably no evolution; its relation shifts slightly upward with time but effectively establishes the local scaling relation as early as $z = 3$. This implies that massive black holes in GIZMO-SIMBA are assembled rapidly in the early Universe (within 2.5 Gyr), tracking the growth of the protocluster halo (T. Di Matteo et al. 2012; K. Inayoshi et al. 2020). TNG-Cluster exhibits a somewhat stronger positive evolution. Its high-redshift progenitors ($z \gtrsim 2$) lie below the local trend by 0.5–1 dex, indicating a delayed growth phase where SMBHs remain undermassive relative to their halos until lower redshifts (R. Weinberger et al. 2018; M. Habouzit et al. 2021). Crucially, these high-redshift offsets are the combined result of subgrid physics and the disparate seeding prescriptions adopted by each model; the choice of seed mass and seeding criteria directly dictates the starting point of the scaling relations before accretion becomes the dominant growth mode.

These differences at $z > 3$ establish a physical distinction between the two models: supply-driven versus feedback-regulated growth. In GIZMO-SIMBA, black hole growth is governed by the gravitational torque-limited accretion model. Unlike thermal-based models, this prescription allows for efficient accretion of cold gas driven by disk instabilities, independent of the halo temperature. Consequently, GIZMO-SIMBA black holes grow rapidly in gas-rich high-$z$ halos (i.e., supply driven). Furthermore, the anisotropic nature of GIZMO-SIMBA's bipolar kinetic jets clears gas from the polar regions but permits intermittent cold inflow along the equatorial plane. This geometry allows the black hole to sustain its mass growth in lockstep with the halo, maintaining the high normalization and early assembly seen in Figure 2.

In contrast, TNG-Cluster employs an unboosted Bondi accretion scheme coupled with isotropic kinetic wind feedback. Since Bondi accretion is highly sensitive to gas sound speed ($\dot{M} \propto c_s^{-3}$), the effective isotropic heating of the halo gas creates a strong negative feedback loop (i.e., the growth becomes feedback regulated). This suppresses early growth in hot, pressurized protoclusters.

In addition, because the Bondi accretion rate in Cluster-TNG scales as $\dot{M} \propto M_{\rm BH}^2\, c_s^{-3}$, the BH growth at the early times is small. At these early times, the mass of the SMBHs is governed by the initial mass seed adopted in the TNG-Cluster. Significant black hole assembly is delayed until the BH mass becomes large enough to overcome the thermal suppression (elevated $c_s$) of the hot protocluster environment. However, this initial slow growth is followed by a rapid one that is able to contribute to the final mass of the SMBHs at later redshifts, almost reaching SMBHs that are as massive as in GIZMO-SIMBA. While the seeding parameters influence the starting point of $M_{\rm BH}$–$M_{200c}$ relation, the post-seeding growth evident in Figure 2 is also governed by feedback regulation. Once the black holes grow sufficiently large, the slope of the $M_{\rm BH}$–$M_{200c}$ relation (also of $M_{\rm BH}$–$M_*$ relation and $M_{\rm BH}$–$\sigma_*$ relation) at fixed $M_{\rm BH}$ in TNG-Cluster is noticeably shallower than the one found in GIZMO-SIMBA. This flattening indicates that TNG-Cluster's strong, isotropic feedback effectively regulates the long-term black hole assembly in the central galaxies that reside in high-mass dark matter halos and cause lower-mass SMBH at a fixed stellar mass/halo mass with respect to GIZMO-SIMBA, which was actually seeded with SMBHs with lower masses.

This monotonic tracking at lower normalizations strongly reflects the theoretical expectations of feedback-regulated baryon lifting, where the central black hole mass is intimately tied to the original binding energy of the halo's baryons (G. M. Voit et al. 2023). That is also why TNG-Cluster requires placing larger BH seeds and applying an early seeding scheme to compensate for this strong early suppression.

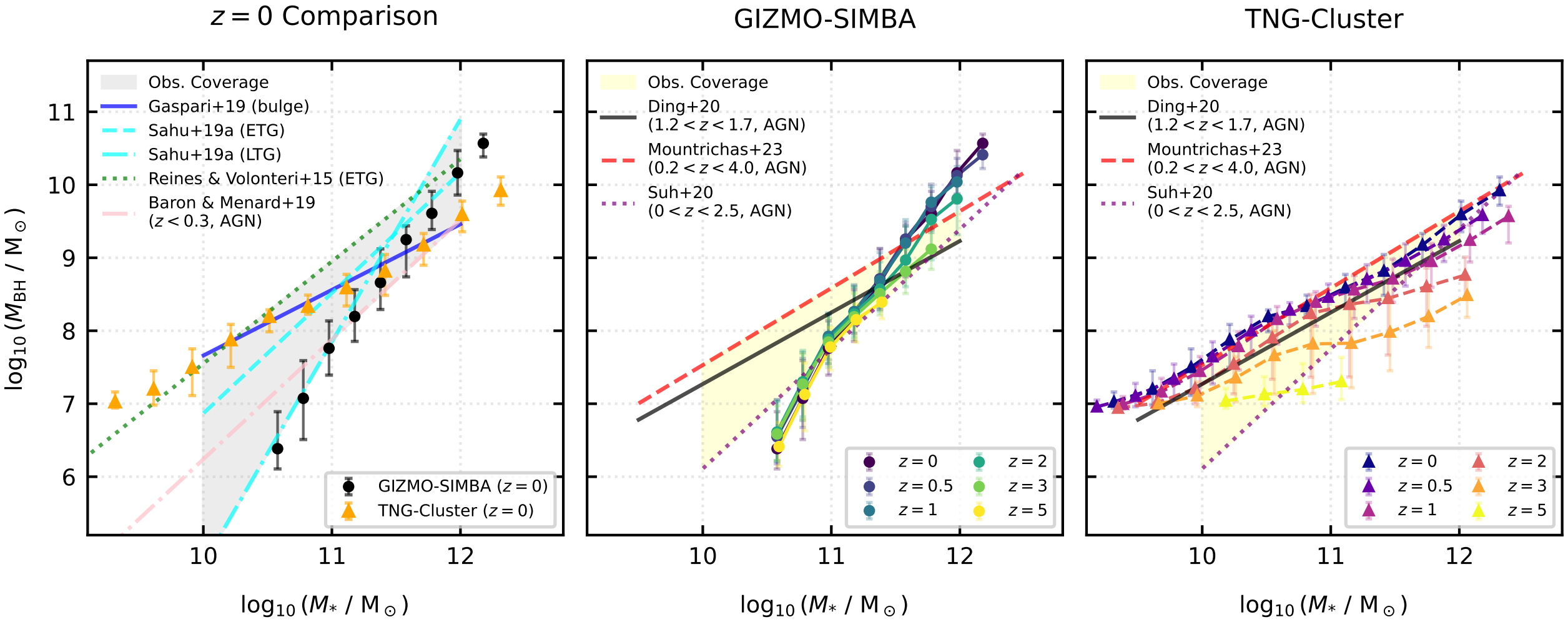

**Figure 3.** The relation between SMBH mass ($M_{\rm BH}$) and stellar mass within 50 kpc ($M_*$) for central galaxies in the GIZMO-SIMBA and TNG-Cluster simulations. The solid and dashed lines connecting the data points represent the median values of the simulated distributions within each mass bin. Error bars denote the 16th–84th percentile range. Left panel: comparison at $z = 0$. GIZMO-SIMBA central galaxies are shown as black circles, while TNG-Cluster central galaxies are plotted as orange triangles. The gray shaded region represents the envelope of local observational scaling relations, encompassing the early-type (ETG) and late-type (LTG) galaxy relations from N. Sahu et al. (2019a), the ETG sample from A. E. Reines & M. Volonteri (2015), along with samples from D. Baron & B. Ménard (2019) and M. Gaspari et al. (2019). Note that all observational relations are stringently restricted to the mass ranges of their respective galaxy samples without extrapolation. Middle and right panels: the redshift evolution of the $M_{\rm BH}$–$M_*$ relation from $z = 5$ to $z = 0$ for GIZMO-SIMBA (middle) and TNG-Cluster (right). For comparison with high-redshift data, we overlay observational constraints from X. Ding et al. (2020), H. Suh et al. (2020), and G. Mountrichas (2023), with the yellow shaded region indicating the coverage of these high-$z$ surveys.

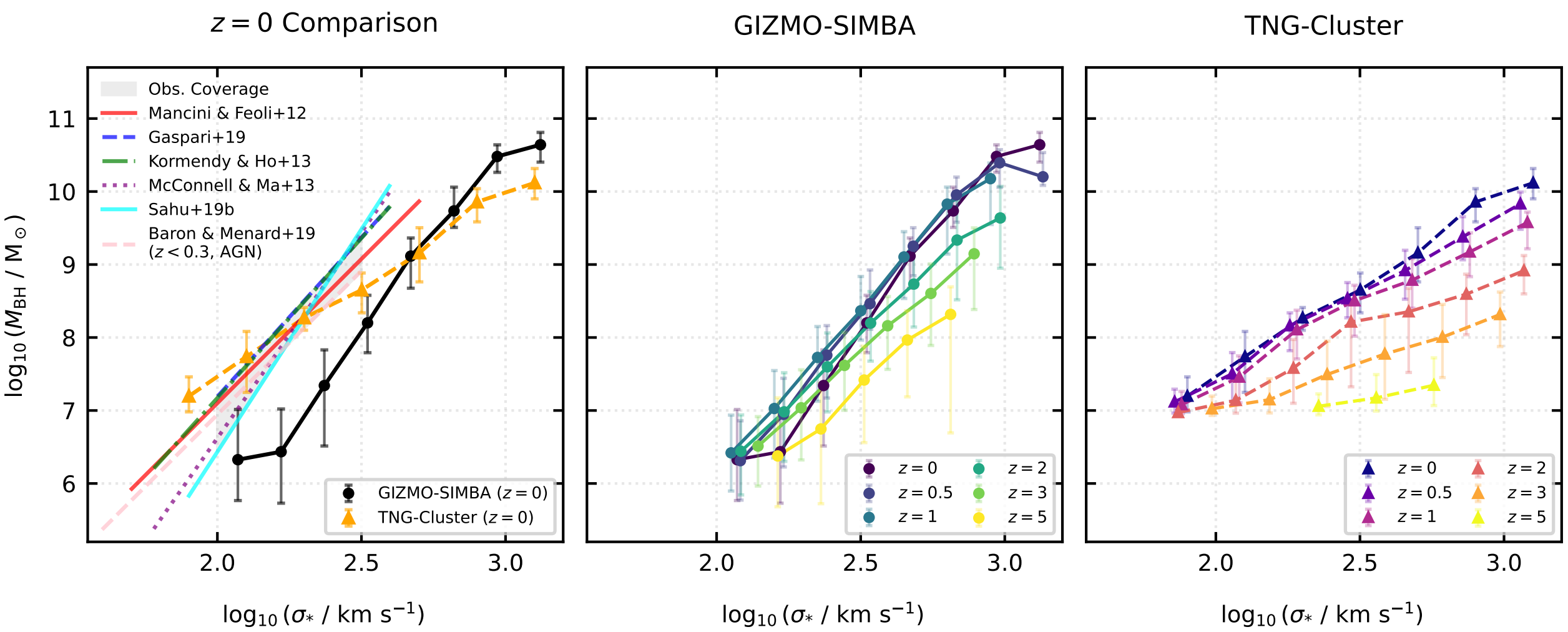

**Figure 4.** The correlation between SMBH mass ($M_{\rm BH}$) and stellar velocity dispersion ($\sigma_*$) for central galaxies in the GIZMO-SIMBA and TNG-Cluster simulations. The solid and dashed lines connecting the data points represent the median values of the simulated distributions within each mass bin. Error bars denote the 16th–84th percentile range. Left panel: comparison of the scaling relations at $z = 0$. GIZMO-SIMBA central galaxies are marked by black circles, while TNG-Cluster central galaxies are shown as orange triangles. The gray shaded region indicates the observational envelope, bounded by the classical bulge-dominated relations of J. Kormendy & L. C. Ho (2013), N. J. McConnell & C.-P. Ma (2013), and L. Mancini & A. Feoli (2012), as well as the relations for broader morphological samples and AGN from N. Sahu et al. (2019b), D. Baron & B. Ménard (2019), and M. Gaspari et al. (2019). Note that all observational relations are stringently restricted to the mass and velocity dispersion ranges of their respective galaxy samples without extrapolation. Middle and right panels: the redshift evolution of the $M_{\rm BH}$–$\sigma_*$ relation from $z = 5$ to $z = 0$ for GIZMO-SIMBA (middle) and TNG-Cluster (right), with data points color-coded by redshift.

### *3.2. Black Hole–Stellar Mass Relation*

In Figure 3 we present the black hole mass ($M_{\rm BH}$)–stellar mass ($M_*$) relationship for central galaxies, comparing the two simulations against a compilation of observed relations that bracket the observational uncertainty. These include early-type galaxy (ETG) and bulge-dominated relations, which typically define the upper envelope (e.g., N. Sahu et al. 2019a, ETG sample; A. E. Reines & M. Volonteri 2015), and late-type galaxy (LTG) or broad-line AGN samples, which extend to lower normalizations characteristic of disk-dominated systems (D. Baron & B. Ménard 2019; N. Sahu et al. 2019a). It is important to note that all overlaid observational relations are stringently restricted to the mass ranges of their respective galaxy samples without any extrapolation.

Consistent with the halo-scale trends we established in Section 3.1, the central galaxies in the two simulations populate distinct regions of the parameter space. For less-massive galaxies ($M_* < 10^{11.5}\,M_\odot$), GIZMO-SIMBA central galaxies at $z = 0$ lie near the lower boundary of the observational band, closely tracking the LTG scaling relations of N. Sahu et al. (2019a). GIZMO-SIMBA also aligns with the ETG relations at the high-mass end ($M_* > 10^{11.5}\,M_\odot$).

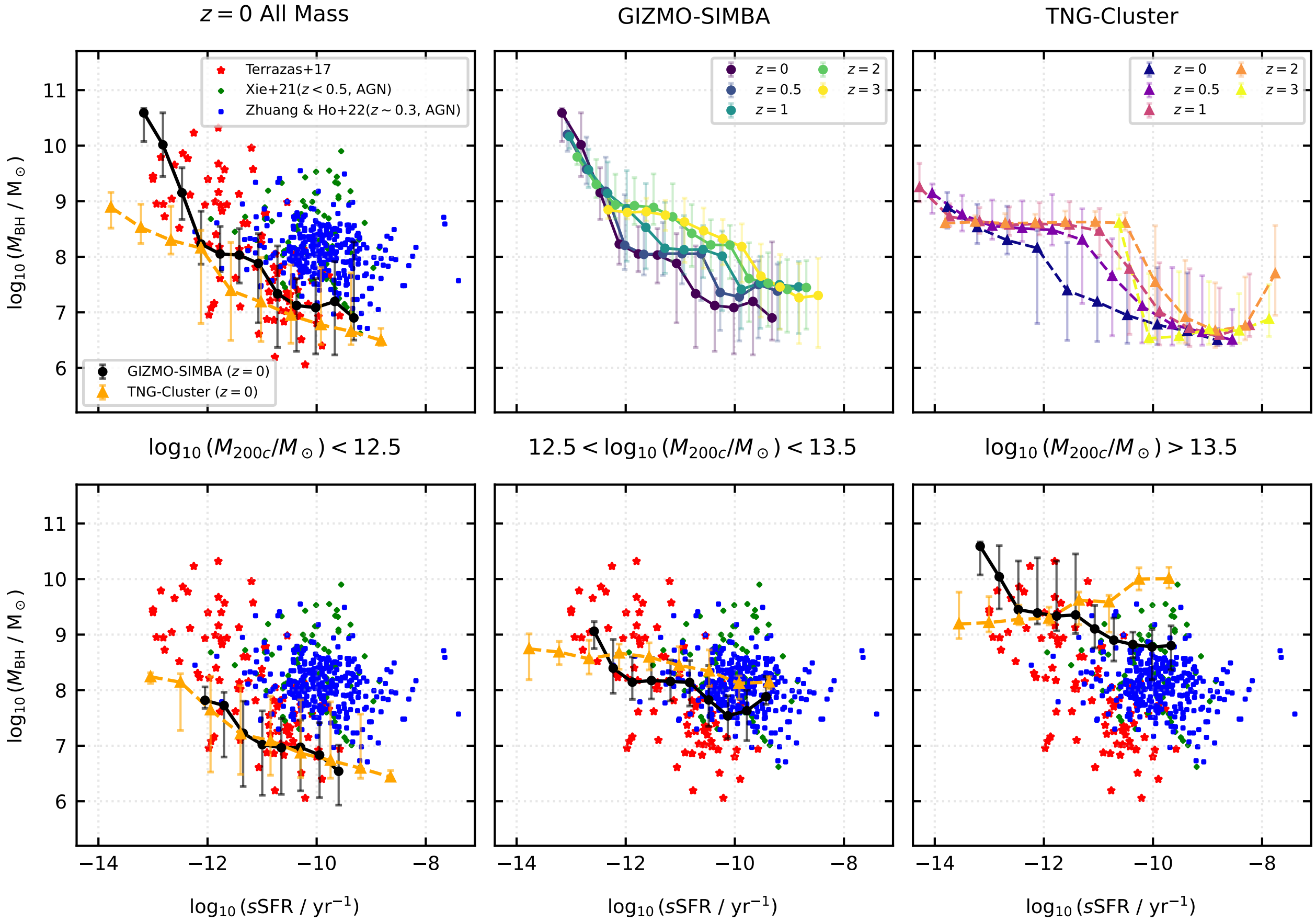

**Figure 5.** The relationship between SMBH mass ($M_{\rm BH}$) and sSFR for central galaxies in the GIZMO-SIMBA and TNG-Cluster simulations. The solid and dashed lines connecting the data points represent the median values of the simulated distributions within each mass bin. Error bars denote the 16th–84th percentile range. Top row: the left panel shows the comparison of the median trends and data distributions at $z = 0$. GIZMO-SIMBA is represented by black circles, and TNG-Cluster by orange triangles. The middle and right panels show the redshift evolution of the $M_{\rm BH}$–sSFR relation from $z = 0$ to $z = 3$ for GIZMO-SIMBA and TNG-Cluster, respectively, color-coded by redshift. The evolution highlights the transition from vigorous star formation at cosmic noon to quiescence in the local Universe. Bottom row: the $z = 0$ scaling relations are divided into three host halo mass bins: $\log_{10}(M_{200c}/M_\odot) < 12.5$ (left), $12.5 < \log_{10}(M_{200c}/M_\odot) < 13.5$ (middle), and $\log_{10}(M_{200c}/M_\odot) > 13.5$ (right). In the $z = 0$ panels (top-left and all bottom panels), observational constraints are overlaid: red stars indicate local central galaxies with dynamically determined black hole masses from B. A. Terrazas et al. (2017), while green diamonds and blue squares represent data from Y. Xie et al. (2021) and M.-Y. Zhuang & L. C. Ho (2022), respectively. Note that these observational data points are stringently restricted to their specific stellar mass ranges: $10^{10}$–$10^{12}\,M_\odot$ for B. A. Terrazas et al. (2017) and M.-Y. Zhuang & L. C. Ho (2022), and $10^{9.5}$–$10^{12}\,M_\odot$ for Y. Xie et al. (2021).

In contrast, TNG-Cluster produces a relation that falls within the observational envelope but behaves differently at different mass scales. Specifically, for objects with $M_* \sim 10^{10}\,M_\odot$, TNG-Cluster aligns with the observations with higher normalization, remaining consistent with the ETG relation of A. E. Reines & M. Volonteri (2015) rather than the median of the literature (which represents also disk-dominated galaxies). At higher masses, its central galaxies show a strong affinity with the relation of M. Gaspari et al. (2019), but the comparison is not fully homogeneous, since the simulation uses stellar mass within 50 kpc, whereas the plotted Gaspari +19 relation is labeled as a bulge relation. This may bias the interpretation for BCGs. Also, M. Gaspari et al. (2019) demonstrated that SMBH mass correlates more fundamentally with the thermodynamics of the host's hot X-ray halo than with classical optical tracers, a finding that supports the CCA framework while disfavoring simple, steady-state Bondi-like fueling.

The redshift evolution panels reveal distinct growth histories that qualitatively mirror the halo-scale trends established in Section 3.1. GIZMO-SIMBA displays a remarkably mild evolution from $z = 5$ to $z = 0$, with its high-redshift progenitors falling within the empirical coverage of high-$z$ AGN surveys. This is consistent with recent JWST findings that suggest little to no significant evolution in the $M_{\rm BH}/M_*$ ratio for galaxies with $\log(M_*/M_\odot) \geqslant 10$ up to $z \sim 4$ (Y. Sun et al. 2024) and that the $M_{\rm BH}$–$M_*$ relation in massive host galaxies remains largely stable from $z \sim 6$ to the present epoch (Y. Sun et al. 2025). In contrast, TNG-Cluster exhibits a steeper evolutionary track; its high-redshift black holes ($z \gtrsim 2$) are significantly undermassive relative to the local relationship and drop below high-$z$ observational constraints before experiencing rapid growth at lower redshifts.

### 3.3. Black Hole–Stellar Velocity Dispersion Relation

Figure 4 illustrates the correlation between black hole mass and stellar velocity dispersion ($\sigma_*$) for central galaxies. The stellar velocity dispersion is a fundamental proxy for the depth of the host galaxy's gravitational potential well.

At $z = 0$, both simulations exhibit a positive correlation. However, they present different normalizations, with GIZMO-SIMBA showing a more significant deviation and falling systematically below the observed relations (e.g., J. Kormendy

& L. C. Ho 2013; N. J. McConnell & C.-P. Ma 2013). TNG-Cluster aligns well with the observations in the median-$\sigma_*$ regime ($\sigma_* \sim$ 100–300 km s$^{-1}$) but shows offsets at the low-$\sigma_*$ end and the high-$\sigma_*$ end. When interpreting this discrepancy, we must consider several factors. First, we note that all of the relations we compare with cover only the region of the observed data points, which is limited to lower masses, and no extrapolations have been used. Second, the offset between GIZMO-SIMBA and these relations should not be interpreted purely as a model failure, but could also be seen as a signature of morphological bias. The observational relations are heavily biased toward classical, highly concentrated bulges. In contrast, our simulated sample encompasses central galaxies with a wider variety of kinematic structures, including systems with rotation-dominated components, which are known to host lower-mass black holes at a fixed $\sigma_*$. Third, the observational relations we employ typically measure the line-of-sight velocity dispersion within a centralized aperture, such as the effective radius ($R_e$) or a central fiber, focusing strictly on the kinematics of the inner bulge. In contrast, the $\sigma_*$ derived from our simulation catalogs represents the global 1D velocity dispersion of all gravitationally bound star particles within the central subhalo. For massive BCGs, these bound stars extend far into the outer envelope, where they begin to trace the deeper, kinematically "hotter" potential of the host galaxy cluster. Integrating over this entire bounded structure systematically inflates the measured $\sigma_*$ relative to central observational apertures. This effect artificially shifts the simulated data points to the right (higher $\sigma_*$) in the $M_{\rm BH}$–$\sigma_*$ plane, which significantly contributes to the simulated populations falling "below" the observed relations. Furthermore, this aperture-driven increase in velocity dispersion is mass-dependent and becomes increasingly noticeable for high-mass halos. As shown in Figure 1, the overall halo and stellar mass distributions of TNG-Cluster sample are systematically lower than those in GIZMO-SIMBA. Consequently, the central galaxies in TNG-Cluster are subjected to a relatively weaker kinematic heating effect from the extended cluster environment. This partially explains why the TNG-Cluster population exhibits a less severe artificial shift and remains relatively closer to the observational $M_{\rm BH}$–$\sigma_*$ relation compared to the more-massive GIZMO-SIMBA sample.

Investigating the high-redshift regime reveals a significant structural evolution. However, it is crucial to preface this with a note of caution regarding observational comparisons. High-redshift samples for the $M_{\rm BH}$–$\sigma_*$ relation are exceedingly rare and suffer from severe selection biases (e.g., the requirement to detect extremely luminous AGN at cosmological distances). Some observational studies tend to suggest a "positive evolution"—namely, more massive black holes at a fixed $\sigma_*$ at higher redshifts, but this is now widely recognized by the astronomical community as primarily an artifact of observational selection effects rather than a true physical evolution (T. R. Lauer et al. 2007; Y. Shen & B. C. Kelly 2010; F. Shankar et al. 2016).

In contrast to observational (which, as we discuss above, are maybe limited), the intrinsic evolution predicted by both models physically shifts downwards in the $M_{\rm BH}$–$\sigma_*$ plane. The two models exhibit qualitatively similar overall trends, with both simulations predicting higher velocity dispersions at high redshifts for a fixed black hole mass. This trend seems to be physically driven by the higher compactness of galaxies in the early Universe (A. van der Wel et al. 2014; S. Wellons et al. 2015; G. Barro et al. 2017). High-$z$ central galaxies reside in denser dark matter halos and have significantly smaller effective radii, leading to deeper, more concentrated potential wells and elevated kinematic dispersions (driven by both rotation and random motions) compared to $z = 0$ galaxies of similar mass. Furthermore, the evolutionary shift in GIZMO-SIMBA is relatively mild, while the evolutionary trend in TNG-Cluster is more apparent.

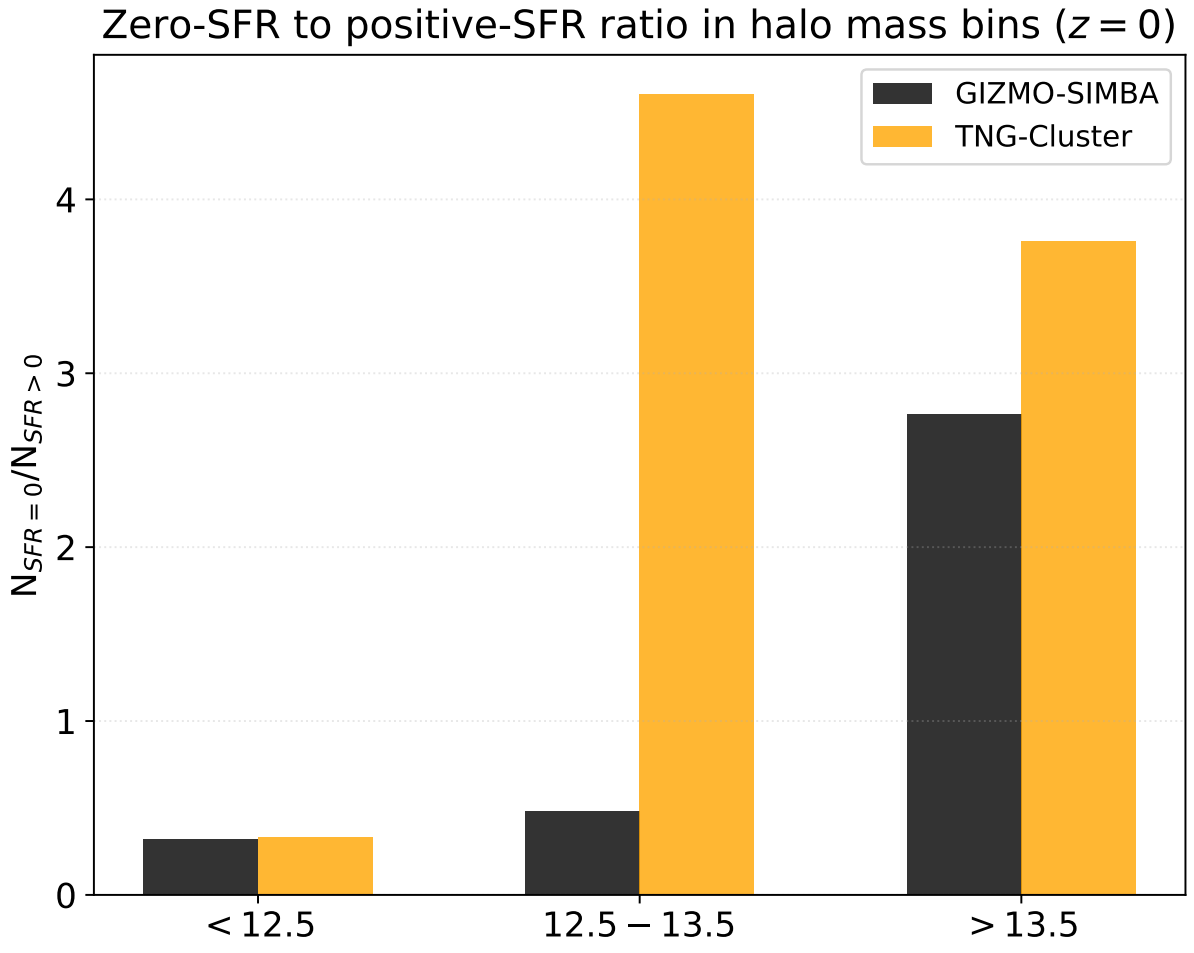

**Figure 6.** Ratio of zero-SFR to positive-SFR central galaxies within three discrete halo mass bins: $\log_{10}(M_{200c}/M_\odot) < 12.5$, 12.5–13.5, and >13.5 at $z = 0$. Black and orange bars represent GIZMO-SIMBA and TNG-Cluster, respectively. The ratio $N_{\rm SFR=0}/N_{\rm SFR>0}$ serves as a diagnostic for the prevalence of fully quenched systems.

### 3.4. Black Hole Mass and Quenching

Figure 5 serves as a crucial diagnostic of the prevention capability of the feedback mechanisms involved. We investigate the connection between black hole mass and the cessation of star formation by plotting the black hole mass-specific SFR for central galaxies at $z = 0$–5. Since we cover a large redshift range, this relation can be a good diagnostic of whether the integrated energy output of the black hole (traced by $M_{\rm BH}$) is indeed a primary driver of galaxy quenching.

Across all panels, both GIZMO-SIMBA and TNG-Cluster agree that central galaxies with more-massive black holes consistently show lower sSFR. This trend qualitatively mirrors the observational results of B. A. Terrazas et al. (2017), who demonstrated that $M_{\rm BH}$ is a more fundamental predictor of quiescence than stellar mass or halo mass.

However, the nature of the quiescent state differs significantly between the two models. At $z = 0$, the GIZMO-SIMBA relation aligns broadly well with the observations. It captures the gradual decline in sSFR for intermediate-mass black holes and the transition to quiescence for massive systems ($M_{\rm BH} \gtrsim 10^8\,M_\odot$). Possibly, this reflects the fact that the anisotropic nature of SIMBA's kinetic jet feedback allows for galaxies to retain a nonzero sSFR, even having massive BHs: while the jets effectively evacuate gas along the polar axis, they allow for intermittent cold-gas inflow along the equatorial plane. This maintenance mode feedback sustains a passive galaxy that is not entirely dead (a "soft quench") (P. N. Best & T. M. Heckman 2012; T. M. Heckman & P. N. Best 2014). Crucially, while a large fraction of massive central galaxies in GIZMO-SIMBA are completely quenched with an exactly zero SFR (as quantified later in Figure 6), a substantial population still maintains a low but widely diverse specific SFR. This aligns with the findings of A. Katsianis et al. (2021b), who noted that a distinct fraction of quenched galaxies in the SIMBA simulations retain nonzero sSFRs.

The TNG-Cluster model predicts significantly lower sSFR for central galaxies at the high BH mass end, creating a sharp cutoff in BH mass. This reflects the high efficiency of the TNG isotropic kinetic wind mode, which acts as an aggressive preventative mechanism by uniformly heating the CGM and shutting down cooling flows entirely. This results in a "hard quench," pushing the sSFR to extreme lower limits near zero (D. Nelson et al. 2019).

The evolutionary panels reveal that the entire $M_{\rm BH}$–sSFR sequence shifts toward the right (higher sSFR) at higher redshifts. At $z \gtrsim 2$, even central galaxies with massive black holes maintain relatively high sSFR, placing them on or near the star-forming main sequence. This behavior is consistent with recent observational studies of quasar host galaxies (e.g., Y. Xie et al. 2021; M.-Y. Zhuang & L. C. Ho 2022), which find that vigorous star formation and rapid black hole growth can coexist in gas-rich environments before feedback fully quenches the galaxy. The decline in sSFR toward $z = 0$ thus reflects the combined effect of a diminishing cosmic gas supply and the cumulative, integrated impact of AGN feedback over time.

To explicitly disentangle the influence of the host halo from the integrated AGN energy on galaxy quenching, the bottom row of Figure 5 divides the $z = 0$ sample into three discrete halo mass bins. This reveals a fascinating, mass-dependent divergence between the two simulation models.

In the lowest halo mass regime ($\log_{10}(M_{\rm 200c}/M_\odot) < 12.5$), the simulated central galaxies naturally populate the lower-mass end of the observational data. Here, the anticorrelation between $M_{\rm BH}$ and sSFR is clearly established. Remarkably, the predictions from GIZMO-SIMBA and TNG-Cluster are nearly identical in this regime. Physically, central galaxies in these lower-mass halos are predominantly gas-rich and star-forming. In both models, the strong, preventative kinetic feedback modes (which act to fully quench galaxies) have not yet dominated. Instead, the black holes and galaxies coevolve via steady gas supplies, regulated by milder feedback channels, resulting in a highly consistent coevolutionary track before catastrophic quenching occurs.

Moving to the intermediate halo mass bin ($12.5 < \log_{10}(M_{\rm 200c}/M_\odot) < 13.5$), the simulations cover the middle-mass range of the observational data. In this regime, both simulations exhibit a significant flattening or weakening of the $M_{\rm BH}$–sSFR relation, almost showing no distinct correlation. Despite this morphological shift, the two models remain highly consistent with one another. This intermediate bin represents a critical transition regime where virial shock heating begins to establish a stable hot CGM (M. S. Calzadilla et al. 2022). The flattening of the relation suggests that during this transitional phase, the complex, messy interplay between diminishing cold-gas accretion, environmental heating, and the onset of strong AGN feedback creates a large scatter. This temporarily decouples the instantaneous SFR from the integrated black hole mass as galaxies traverse the "green valley."

Finally, in the most massive halo environment ($\log_{10}(M_{\rm 200c}/M_\odot) > 13.5$), overlapping with the high-mass observational sample, a divergence emerges, and the two models yield opposite results. For GIZMO-SIMBA, the strict anticorrelation between $M_{\rm BH}$ and sSFR persists, which aligns well with the downward trend of the observational constraints in the higher-mass data range. Conversely, in TNG-Cluster, the anticorrelation completely disappears, flattening out and even hinting at a slight positive correlation for the most massive black holes. This again highlights the different quenching scenarios ("soft quench" versus "hard quench") in two simulations that we discussed before.

To make the above statement clearer, we perform an investigation, focusing of the missing zero-SFR galaxies. As presented in Figure 6, while both simulations show an increasing fraction of quenched galaxies at higher halo mass, GIZMO-SIMBA (black bars) maintains a lower ratio across all bins, which is consistent with the "soft quench" scenario. In contrast, TNG-Cluster (orange bars) shows a dramatic increase in the ratio for intermediate-mass and massive halos, reflecting the "hard quench" nature of its isotropic kinetic wind feedback, which more effectively eradicates cold-gas reservoirs.

### *3.5. AGN-regulated Gas Depletion*

To understand the physical origin of the suppression of star formation observed in Figure 5, we examine the evolution of the baryon reservoirs in Figure 7. We plot the simulated black hole masses against three gas-to-stellar mass ratios for central galaxies: total gas ($M_{\rm gas}/M_*$), molecular gas ($M_{\rm H_2}/M_*$), and atomic hydrogen ($M_{\rm H\,I}/M_*$).

Across all three phases of the ISM, both simulations reveal a strong anticorrelation: central galaxies hosting more-massive black holes systematically possess lower gas fractions (S. Appleby et al. 2020; R. Davé et al. 2020). This trend suggests that the quenching observed in the previous section is consistent with the physical removal or heating of the gas supply via ejective and preventative feedback. However, the morphological features of this depletion differ drastically between the two models. Rather than mere gas removal, these processes are better framed as the regulation of a multiphase cycle, where feedback dictates the transition between the hot atmosphere, warm filaments, and the cold ISM (M. Gaspari et al. 2018, 2020). Such predicted behaviors align with observational anchors that demonstrate a complex interplay between gas cooling and AGN feedback in massive ellipticals (P. Temi et al. 2018).

In GIZMO-SIMBA, as presented in W. Cui et al. (2021) and later confirmed in W. Ma et al. (2025), the kinetic jet mode feedback—rather than the quasar mode—is primarily responsible for quenching galaxies. This jet mode feedback mostly peaks around $z \sim 2$. Subsequently, there is a gradual but clear decrease in the gas mass fractions (especially $H_2$, which SIMBA uses for its star formation model). This decreasing trend has been thoroughly discussed in L. J. Tacconi et al. (2020) from the observational point of view. Conversely, TNG-Cluster exhibits a steep drop-off in all gas fractions once the black hole mass exceeds $\sim 10^8\,M_\odot$. This efficient gas depletion reflects the high impact of TNG's isotropic kinetic wind mode, which effectively heats and redistributes the gas reservoir, resulting in rapid quenching of star formation.

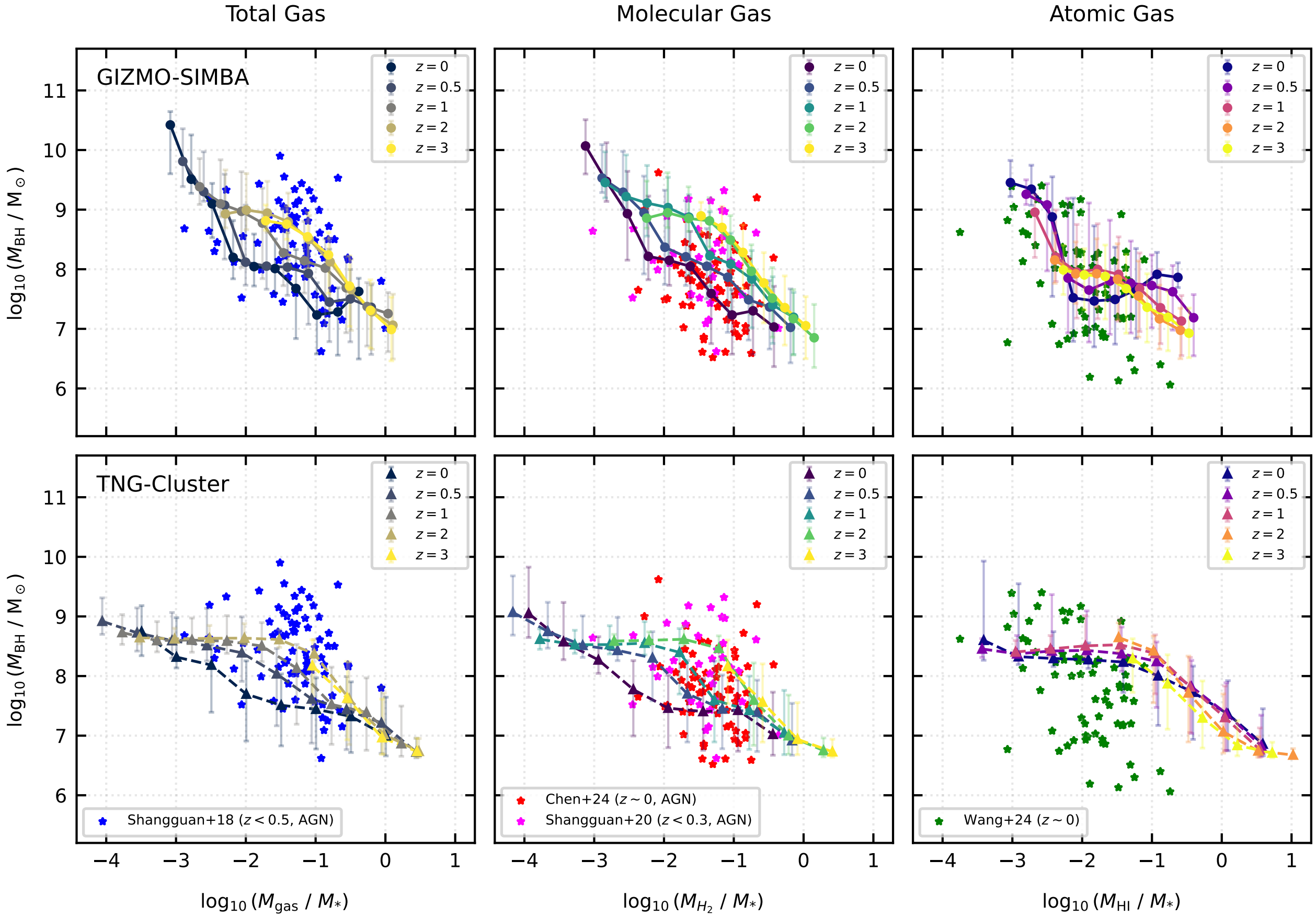

**Figure 7.** The relationship between SMBH mass ($M_{BH}$) and galaxy gas content for central galaxies in the GIZMO-SIMBA (top row, circles) and TNG-Cluster (bottom row, triangles) simulations. The solid and dashed lines connecting the data points represent the median values of the simulated distributions within each mass bin. Error bars denote the 16th–84th percentile range. The three rows display the dependence of $M_{BH}$ on (left) total gas fraction ($M_{gas}/M_*$), (middle) molecular gas fraction ($M_{H_2}/M_*$), and (right) atomic gas fraction ($M_{HI}/M_*$). Different colored symbols and lines represent the data distribution and median trends at different redshifts, evolving from $z = 3$ to $z = 0$. To benchmark the gas predictions, we overlay observational measurements of host galaxies: J. Shangguan et al. (2018) (blue stars; total gas in quasars), Y. Chen et al. (2024) (red stars; molecular gas in AGNs), J. Shangguan et al. (2020) (orange stars; molecular gas in quasars), and T. Wang et al. (2024) (green stars; atomic gas in central galaxies). Note that all observational samples are subject to detection limits and are stringently restricted to their specific stellar mass ranges: $10^{10}$–$10^{12}\,M_\odot$ for J. Shangguan et al. (2018), Y. Chen et al. (2024), and T. Wang et al. (2024); and $10^{9.5}$–$10^{12}\,M_\odot$ for J. Shangguan et al. (2020).

In the middle panels of Figure 7, we present the BH mass–molecular gas relation; both simulations broadly overlap with the observational data points from Y. Chen et al. (2024) and SJ. Shangguan et al. (2020). However, the observed data show a large scatter and a somewhat shallower dependence compared to the simulations. This discrepancy likely arises from selection effects: observational samples are often biased toward detectable, gas-rich systems (often hosting active AGNs) and may miss the population of fully quenched, gas-depleted galaxies that populate the quiescent tail of our simulation (A. Saintonge & B. Catinella 2022).

The atomic gas relation (right panels) follows a clear declining sequence, qualitatively confirming the recent findings of T. Wang et al. (2024), who showed that the HI gas fraction is fundamentally anticorrelated with black hole mass. Both simulations support this physical picture, yet they bracket the observational trend. GIZMO-SIMBA shows a shallower slope compared to the steep decline observed by T. Wang et al. (2024). This is likely a consequence of its anisotropic jet geometry, which expels gas perpendicular to the disk but allows residual neutral gas to survive in the equatorial plane. In contrast, when black hole mass does not reach $\sim 10^8\,M_\odot$, TNG-Cluster shows no trend, as the feedback is not yet efficient enough to heat the gas. However, once the black hole mass crosses this critical threshold, TNG-Cluster predicts a steep drop in the HI fraction. Last, we note that TNG-Cluster's galaxies with $M_{BH} > 10^8\,M_\odot$ are more abundant in the atomic gas than the observed sample and GIZMO-SIMBA.

Regarding redshift evolution, we find that while the total and molecular gas fractions in both simulations evolve strongly (reflecting cosmic downsizing), their $M_{BH} - M_{HI}/M_*$ relations show much weaker evolution. This suggests that in both simulations, while the total baryon budget shrinks, the relative abundance of atomic hydrogen in surviving disks remains regulated by a steady balance between cooling, star formation, and feedback across cosmic time.

### *3.6. Black Hole Fueling versus Stellar Mass*

In Figure 8 we present the instantaneous growth rates that possibly drive the integrated relations discussed in Section 3.2. We examine the dependence of BHAR on stellar mass for central galaxies, separating the population into star-forming (SF) and quenched (Q) systems using a redshift-dependent sSFR threshold ($sSFR_{lim} = 10^{(-1.8+0.3z)}$ Gyr$^{-1}$; R. Davé et al. 2019). This separation allows us to isolate the fueling efficiency in different types of galaxies.

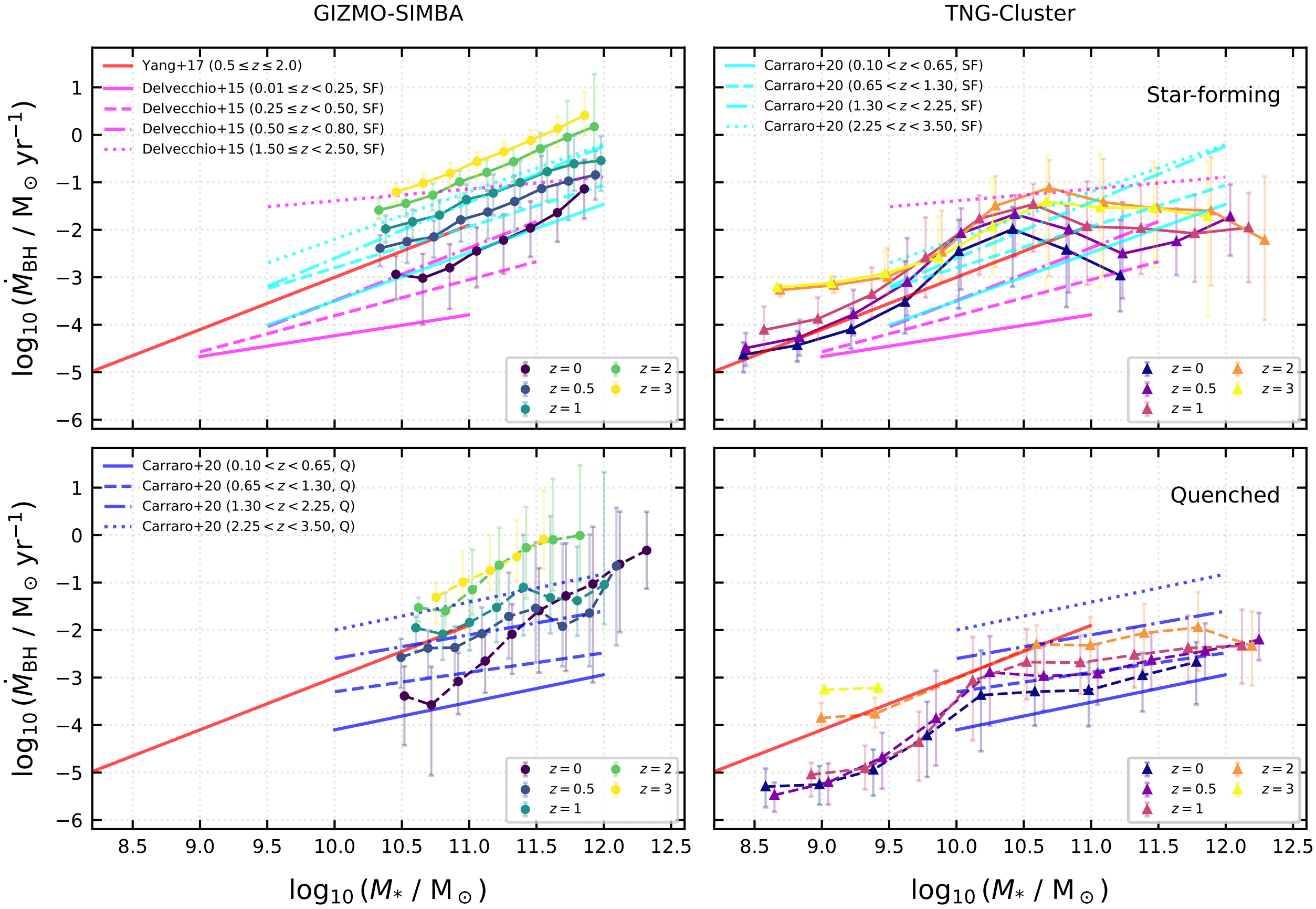

**Figure 8.** The dependence of black hole accretion rate (BHAR) on stellar mass ($M_*$) for central galaxies in the GIZMO-SIMBA (left column, circles) and TNG-Cluster (right column, triangles) simulations. The solid and dashed lines connecting the data points represent the median values of the simulated distributions within each mass bin. Error bars denote the 16th–84th percentile range. The galaxy populations are separated into star-forming (top row) and quenched (bottom row) systems. The classification in the simulations is based on the sSFR threshold $\mathrm{sSFR_{lim}} = 10^{(-1.8+0.3z)}\ \mathrm{Gyr}^{-1}$ (R. Davé et al. 2019). The colored lines trace the redshift evolution of the median BHAR–$M_*$ relations from $z = 3$ to $z = 0$. To contextualize the simulation predictions, we overlay observational constraints for star-forming and quenched galaxies from G. Yang et al. (2017), I. Delvecchio et al. (2015), and R. Carraro et al. (2020). Note that all observational data and fitted relations are stringently restricted to the stellar mass ranges of their respective samples without any extrapolation.

Before comparing with the empirical data, it is crucial to discuss the differences in how star-forming and quenched galaxies are classified. While our simulations utilize a strict sSFR threshold, observational studies (such as I. Delvecchio et al. 2015 and R. Carraro et al. 2020) typically differentiate these populations using color–color diagrams (e.g., UVJ cuts) or by measuring their distance from the empirical star-forming main sequence. These varying criteria can naturally introduce some systematic offsets or scatter, particularly for galaxies caught in the transition phase. However, the fundamental dichotomy remains intact: both simulated sSFR cuts and empirical color/main-sequence cuts broadly trace similar underlying bimodal macroscopic populations. The exact boundaries may differ, but the qualitative trends that show the fueling efficiency of actively star-forming galaxies are robust and different than the ones of quiescent systems.

For the star-forming central galaxies (top row), both simulations exhibit a positive correlation between BHAR and stellar mass, with normalizations declining steadily from $z = 3$ to $z = 0$ due to cosmic downsizing. GIZMO-SIMBA exhibits a tight, roughly linear correlation across all redshifts that aligns remarkably well with the slopes and normalizations of the observed relations from G. Yang et al. (2017) and R. Carraro et al. (2020). This good agreement suggests that in gas-rich disks, black hole growth is firmly governed by the global mass assembly of the host (i.e., coevolutionary growth). TNG-Cluster predicts a broadly similar slope but displays noticeable nonlinear inflections or "wiggles" at intermediate masses before flattening out at the highest masses.

For the quenched population (bottom row), a striking physical divergence emerges, perfectly mirroring the distinct quenching mechanisms discussed in Section 3.4. GIZMO-SIMBA predicts sustained, unexpectedly high-level of accretion rates in massive quenched galaxies ($M_* \gtrsim 10^{10.5}\,M_\odot$), with the median BHAR often curving upward to match or exceed the rates found even in some star-forming systems. Consequently, GIZMO-SIMBA significantly overpredicts the BHAR compared to the quenched observational constraints from R. Carraro et al. (2020). This implies that massive black holes in GIZMO-SIMBA continue to accrete efficiently (via maintenance mode from residual equatorial gas) even after their host galaxies have ceased forming stars. We also tested the BHAR evolutionary history of massive BHs in GIZMO-SIMBA simulation, and Figure 9 shows a representative one. The result is that, at high redshifts ($z \gtrsim 2.5$), BH growth is primarily dominated by the efficient accretion of cold gas via gravitational torques. As the host halo grows and virializes ($z \lesssim 2$), the fueling mechanism transitions to be dominated by Bondi accretion from the hot ICM, allowing the SMBH to maintain growth even after the host galaxy's star formation is quenched. This result confirms the above statement of elevated BHAR in quenched galaxies for GIZMO-SIMBA. In contrast, TNG-Cluster shows a obvious suppression of BHAR in massive quiescent galaxies. This hard starvation behavior aligns much better qualitatively with the low BHARs observed in the R. Carraro et al. (2020) quenched sample, reflecting the highly effective isotropic clearance of the gas reservoir in the TNG subgrid model.

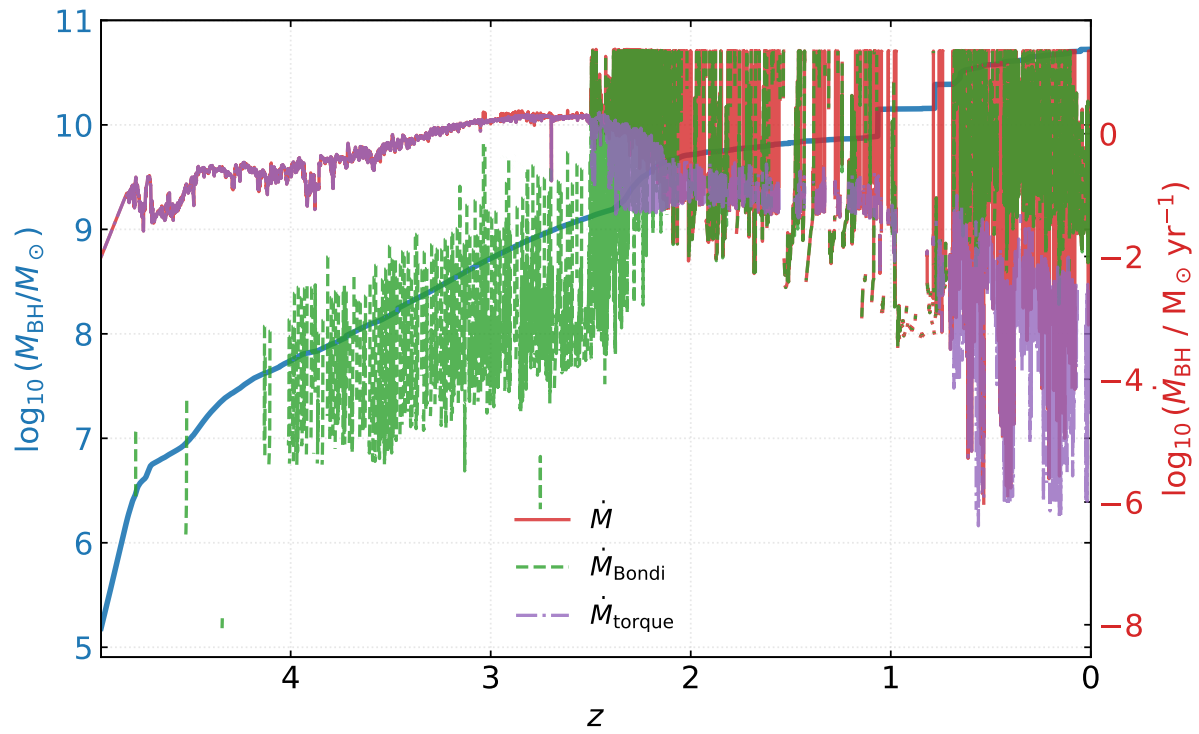

**Figure 9.** Accumulative $M_{\rm BH}$, $\dot{M}_{\rm BH}$, $\dot{M}_{\rm Bondi}$, and $\dot{M}_{\rm torque}$ of a representative central SMBH in the GIZMO-SIMBA simulation. The blue solid line tracks the growth of the SMBH mass ($\log_{10}M_{\rm BH}$, left $y$-axis) from its initial seeding at high redshift to $z = 0$. The right $y$-axis displays the BHAR ($\log_{10}\dot{M}_{\rm BH}$ in $M_\odot$ yr$^{-1}$). The total accretion rate ($\dot{M}$, red line) is decomposed into two distinct modes: the cold-gas torque-limited accretion ($\dot{M}_{\rm torque}$, purple line) and the hot-gas Bondi accretion ($\dot{M}_{\rm Bondi}$, green dashed line).

### *3.7. BHAR–SFR Relation*

Figure 10 examines the coevolution of black holes and central galaxies in the BHAR–SFR plane. This diagnostic is critical for distinguishing whether AGN activity strictly tracks the availability of cold gas (which fuels star formation) or proceeds via a distinct channel in passive galaxies.

For the star-forming population (top row), both simulations exhibit a strong, positive correlation between BHAR and SFR, broadly covering the regions mapped by observational constraints. However, GIZMO-SIMBA reveals a distinct, systematic redshift evolution that warrants physical attention: at a fixed SFR, the simulation predicts a noticeably higher accretion rate at high redshifts compared to lower redshifts. Fundamentally, GIZMO-SIMBA suggests that at the same SFR, there is more cold gas concentrated around the central black hole at high-$z$ (driving a higher accretion rate), while at low-$z$, the remaining cold gas is distributed farther outside the black hole, mostly located in or around the galactic disks (driving lower central accretion but sustaining SFR). This is consistent with an inside-out quenching interpretation (G. R. Tremblay et al. 2018). It is important to note that observational evidence sometimes points to the opposite evolutionary trend, suggesting a potential tension and maybe pointing to the direction that further refinement or mass-dependent calibration have to undergo for the subgrid accretion model.

For the quenched population (bottom row), the simulations provide a definitive test of the feedback and fueling geometry. GIZMO-SIMBA displays a distinct high-BHAR behavior where quenched central galaxies maintain elevated accretion rates relative to their low SFR. Remarkably, this prediction aligns well with the observational results of M. McDonald et al. (2021) and G. Yang et al. (2019), who found that the ratio of BHAR to SFR is significantly enhanced in massive cluster ellipticals and bulge-dominated systems compared to the field.

The physical driver behind this behavior in the simulation is a fundamental shift in the fueling mechanisms. The underlying factor is that after $z \sim 2$, the accretion in these massive halos in GIZMO-SIMBA becomes dominated by Bondi accretion fueled by the abundant hot intracluster and circumgalactic gas. Consequently, the AGN operates in a sustained hot-halo feeding phase: the anisotropic jet feedback effectively expels gas along the polar direction (preventing star formation), while the deep potential well continuously feeds the black hole via hot-gas accretion, sustaining the BHAR (M. Gaspari et al. 2015; G. R. Tremblay et al. 2016).

Similarly, TNG-Cluster reveals a distinct decoupling between black hole accretion and star formation for the quenched regime, characterized by a noticeably flatter overall slope. Rather than a severe starvation of the black hole, the simulated BHAR in TNG-Cluster remains elevated even at the extreme low-SFR end. This behavior results in a significantly enhanced BHAR-to-SFR ratio that also qualitatively agrees with the observational constraints from massive ellipticals (e.g., M. McDonald et al. 2021). Physically, this indicates that while the highly efficient isotropic kinetic wind in TNG-Cluster completely suppresses star formation by heating and evacuating the cold ISM, the central black hole continues to be persistently fueled and grow. Because the TNG subgrid model employs a Bondi-based accretion scheme, the massive black holes in these quenched central galaxies can still efficiently accrete from the dense, hot, pressure-supported CGM. This continuous hot-halo feeding mechanism sustains ongoing AGN activity long after the cold-gas supply is depleted, producing the shallow dependence of BHAR on SFR observed in the quiescent tail.

## 4. Caveats and Discussion

Recent observational studies have provided connections between SMBHs to other properties of galaxies like their cold-gas contents. Large surveys of atomic and molecular gas have revealed strong anticorrelations between SMBH mass and gas fractions, particularly in massive galaxies, implying that AGN feedback plays a key role in depleting or preventing the replenishment of cold-gas reservoirs (A. Saintonge et al. 2017; Y. Chen et al. 2024; T. Wang et al. 2024). These results suggest that black hole mass may be a more fundamental predictor of gas depletion and quenching than stellar mass or halo mass alone, and they support a halo-baryon regulated picture, providing a stringent observational test for theoretical models (M. Gaspari et al. 2019). Indeed, W. Cui et al. (2021) clearly shows the strong linear correlation between the AGN feedback with the galaxy quenching time in SIMBA simulation, and recent analyses utilizing machine learning on cosmological simulations have explicitly identified black hole mass as the most critical parameter determining the quiescent state of central galaxies (B. A. Terrazas et al. 2020; A. F. L. Bluck et al. 2023). Furthermore, while instantaneous AGN activity may coincide with gas-rich phases (M. J. Cowley et al. 2018; S. R. Ward et al. 2022), the cumulative energy injection traced by black hole mass is thought to be the primary driver of long-term gas depletion. Using GIZMO-SIMBA and TNG-Cluster, we explored how SMBHs are tied to other galaxy properties.

However, it is worth noting that when interpreting the physical differences discussed in the above sections, it is necessary to address the potential impact of numerical resolution. The two simulation suites analyzed in this work employ different hydrodynamical solvers and resolution settings, as detailed in their respective introductory papers: W. Cui et al. (2022) for GIZMO-SIMBA and D. Nelson et al. (2024) for TNG-Cluster. Generally, the TNG-Cluster simulations offer a mass resolution comparable to the TNG300 tiers, while the standard GIZMO-SIMBA runs within The Three Hundred project are designed to efficiently sample a large volume of clusters at a resolution sufficient to capture global scaling relations. Black hole accretion models can be sensitive to these resolution differences. Specifically, the Bondi accretion prescription is theoretically resolution-dependent, as the Bondi radius is often unresolved in cosmological volumes. However, as discussed in D. Nelson et al. (2024), the TNG-Cluster model inherits the robust numerical convergence.

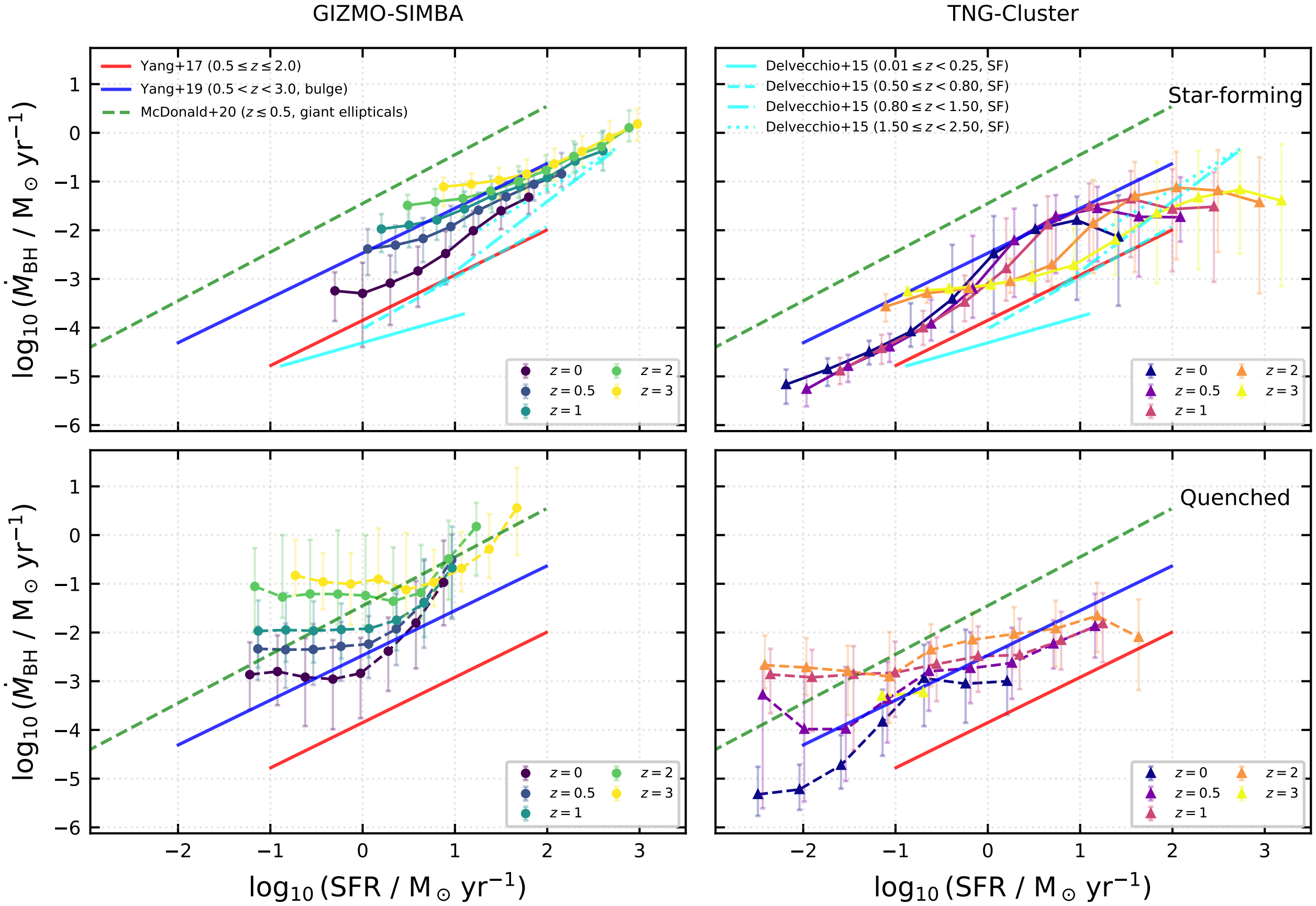

**Figure 10.** The correlation between BHAR and SFR for central galaxies in the GIZMO-SIMBA (left column, circles) and TNG-Cluster (right column, triangles) simulations. The solid and dashed lines connecting the data points represent the median values of the simulated distributions within each mass bin. Error bars denote the 16th–84th percentile range. The populations are separated into star-forming (top row) and quenched (bottom row) systems based on the sSFR threshold: $\mathrm{sSFR_{lim}} = 10^{(-1.8+0.3z)}\ \mathrm{Gyr}^{-1}$ (R. Davé et al. 2019). The colored lines trace the redshift evolution of the median relations from $z = 3$ to $z = 0$. Observational scaling relations are overlaid for comparison: the constraints for galaxies from G. Yang et al. (2017) and for star-forming galaxies from I. Delvecchio et al. (2015), as well as measurements for bulge-dominated and giant elliptical systems from G. Yang et al. (2019) and M. McDonald et al. (2021). Note that all observational relations and data points are stringently restricted to the property ranges of their respective galaxy samples without any extrapolation.

In addition, the torque-limited accretion model used in GIZMO-SIMBA is formulated based on integrated gas properties within the accretion kernel and large-scale disk instabilities. W. Cui et al. (2022) demonstrated that this subgrid model produces consistent black hole populations even at the resolution of The Three Hundred project, as it is comparatively less sensitive to local peak density variations than thermal-based capture models. Crucially, we argue that the divergent evolutionary tracks identified in this work are driven fundamentally by the functional forms of the subgrid physics rather than numerical resolution. Although subgrid parameters (e.g., feedback efficiency normalizations or velocity caps) are recalibrated at different resolution levels to maintain macroscopic energy coupling, the fundamental physical mechanisms—such as SIMBA's torque-limited accretion and bipolar kinetic jets versus TNG's Bondi accretion and isotropic thermal/kinetic feedback—remain conceptually unaltered. The distinct functional forms of the different subgrid models dictate the conditions and energy coupling over cosmic time, thereby shaping the distinct evolutionary trajectories of the black hole–galaxy scaling relations across different redshifts in the two simulations. This interpretation is supported by cross-simulation comparisons like M. Habouzit et al. (2021), which compared SIMBA, TNG, and EAGLE physics in uniform volumes with matched resolutions and reported qualitative differences consistent with our findings. In addition, we think that the evolutionary tracks are more subgrid-physics-dominated rather than the resolution, because these clusters are generally with over 1 million particles, which is way above the resolution. Therefore, although higher resolution can refine the details of small-scale gas clumpiness, the core differences we show arise from a robust physical difference in the theoretical models themselves.

Beyond numerical resolution, we must also emphasize a fundamental caveat regarding the direct comparison between our simulation results and observational data. In this study, galaxy properties such as velocity dispersion, SFR, and BHAR are derived directly from the intrinsic, instantaneous physical features of the simulation particles and cells. This approach is not strictly equivalent to observational measurements, as we did not generate mock observations to account for observational biases, projection effects, or specific instrument limitations. Furthermore, observational methodologies for measuring SFR and BHAR rely on various tracers and scaling relations that carry their own distinct, and sometimes substantial, systematic uncertainties. For instance, M. McDonald et al. (2021) estimated BHAR in giant elliptical galaxies by calculating the mechanical power of AGN-driven X-ray cavities. Such approaches inherently yield BHAR values with large uncertainties due to underlying assumptions regarding cavity volumes, buoyancy timescales, and energy conversion efficiencies. Consequently, when confronting our simulated scaling relations with observational data, a certain degree of inconsistency is to be expected.

## 5. Conclusions

In this work, we present a systematic analysis of the coevolution of SMBHs and their host central galaxies within the extreme environments of galaxy clusters. To do so, we employ two state-of-the-art zoomed-in simulation suites with distinct philosophies: GIZMO-SIMBA and TNG-Cluster. We investigate how black hole masses relate to host halo mass, stellar mass, gas content, and star formation activity, and we explore the impact of the distinct subgrid prescriptions implemented in the two models. Our conclusions can be summarized as follows:

1. GIZMO-SIMBA represents a supply-driven growth model. By employing a gravitational torque-limited accretion prescription that depends primarily on cold-gas kinematics and disk instabilities rather than thermal pressure, GIZMO-SIMBA allows black holes to grow efficiently in the gas-rich environments of high-redshift protoclusters. This insensitivity to the thermal state of the halo enables the black hole to assemble rapidly closely following the dark matter halo growth. The $M_{\rm BH}$–$M_{200c}$ relation is already set at $z = 5$ (very early epochs), and since then has a very mild redshift evolution (Section 3.1). On the other hand, TNG-Cluster represents a feedback-regulated growth model. Relying on Bondi accretion, which is highly sensitive to the local sound speed, combined with effective isotropic thermal and kinetic feedback, the TNG-Cluster model strongly suppresses black hole growth in early times at hot, and pressurized halos. Significant SMBH assembly is effectively delayed until the host potential well deepens sufficiently to confine the heated gas, leading to a distinct evolutionary phase with higher growth rates at later times and an alignment with the theoretical expectations of feedback-regulated baryon lifting.
2. Both paradigms find support in different observational studies. While both simulations produce $M_{\rm BH}$–$M_*$ relations that fall within the broad observational envelope at $z = 0$, they populate distinct normalizations and morphological sequences (Section 3.2). GIZMO-SIMBA aligns with the relations dominated by LTGs at the low-mass end and classical ETGs at the high-mass end. In contrast, TNG-Cluster distinctly aligns with the relations provided by ETG population at the lower stellar mass regime before converging with the massive bulge relations.
3. We find that in both simulations, black hole mass is a fundamental regulator of cold-gas content. A key finding of our study is the decisive role of feedback geometry and accretion channel switching in regulating galaxy quenching and the baryon cycle.
   a. In TNG-Cluster, the isotropic kinetic wind feedback is consistent with an efficient depletion of cold-gas reservoirs and a subsequent "hard quench" of star formation (Sections 3.4 and 3.5). By uniformly heating the CGM and ICM, this mechanism effectively thermalizes the halo and shuts down cold cooling flows. However, rather than starving the black hole, the sustained BHAR in TNG suggests a decoupled quiescent phase where the central black hole potentially continues to feed from the dense, hot, pressure-supported CGM (Section 3.7).
   b. In contrast, the anisotropic bipolar jets in GIZMO-SIMBA point toward a softer, inside-out quenching process. While these jets efficiently evacuate gas along the polar axis, their directional nature permits intermittent inflows or the survival of residual cold gas along the equatorial plane.
4. As halos grow massive after $z \sim 2$, black hole accretion in GIZMO-SIMBA transitions to become heavily dominated by Bondi accretion from the abundant hot intracluster gas. Both mechanisms (TNG's potential hot-halo feeding and SIMBA's combination of residual cold gas with hot Bondi accretion) are consistent with a state where massive central galaxies remain globally quiescent yet retain fueling channels sufficient to power the central engine at rates significantly higher than their residual star formation would imply (Sections 3.6 and 3.7). This provides a plausible theoretical explanation for the radio-mode AGN observed in cluster centers, which remain active despite the quiescence of their hosts.

Future observations of high-redshift protoclusters with JWST and Euclid, particularly those constraining the black hole-to-galaxy mass ratios and gas spatial distributions at $z > 2$, will be essential to break the degeneracy between these two theoretical paradigms. In addition, improvements in both resolution and subgrid physics will yield more physically motivated simulations that better align with the full range (both in terms of redshifts and masses) of observational studies.

## Acknowledgments

We thank the anonymous reviewer for providing comments and suggestions that have significantly improved our work. This work is supported by the National SKA Program of China (2025SKA0150104). This work has been made possible by the "The Three Hundred" collaboration.[7] A.K. has been supported by the 100 talent program of the Sun Yat-sen University, and the Guangdong Basic and the Applied Basic Research Foundation with No. 2025A1515012670. W.C. gratefully thanks Comunidad de Madrid for the Atracción de Talento fellowship No. 2020-T1/TIC19882 and Agencia Estatal de Investigación (AEI) for the Consolidación Investigadora Grant CNS2024-154838. He further acknowledges the Project PID2024-156100NB-C21 financed by MICIU/AEI/10.13039/501100011033/FEDER, EU and ERC: HORIZON-TMA-MSCA-SE for supporting the LACEGAL-III (Latin American Chinese European Galaxy Formation Network) project with grant No. 101086388 and the science research grants from the China Manned Space Project. M.G. acknowledges support from the ERC Consolidator Grant *Black Hole Weather* (101086804). J.S. is supported by the China Manned Space Program (CMS-CSST-2025-A09) and the National Natural Science Foundation of China (NSFC; grant No. 12573014).

[7] https://www.nottingham.ac.uk/astronomy/The300/index.php

*Software:* AHF (S. R. Knollmann & A. Knebe (2009), http://popia.ft.uam.es/AHF), CAESAR (https://github.com/dnarayanan/caesar).

## ORCID iDs

Weiguang Cui https://orcid.org/0000-0002-2113-4863
Romeel Davé https://orcid.org/0000-0003-2842-9434
Massimo Gaspari https://orcid.org/0000-0003-2754-9258
Weishan Zhu https://orcid.org/0000-0002-1189-2855
Xiaohu Yang https://orcid.org/0000-0003-3997-4606